\documentclass[aps,twocolumn,reprint]{revtex4-2}

\usepackage[latin9]{inputenc}

\usepackage[normalem]{ulem}
\usepackage{color}
\usepackage{mathtools}

\usepackage{graphicx}
\usepackage{hyperref}
\usepackage{xfrac}

\usepackage{graphicx}
\usepackage{float}
\restylefloat{table}

\usepackage{amsmath,amssymb,amsxtra,amstext,amsfonts,dsfont}
\usepackage{multirow}
\usepackage{array}
\usepackage{upgreek}
\usepackage{mathrsfs}

\usepackage[squaren]{SIunits}

\usepackage{soul,xcolor}
\setstcolor{red}

\newcommand{\bra}[1]{\ensuremath{\langle{#1}|}}
\newcommand{\ket}[1]{\ensuremath{|{#1}\rangle}}

\newcommand{\abs}[1]{\ensuremath{\left|{#1}\right|}}
\newcommand{\oa}{\omega_a}
\newcommand{\oq}{\omega_q}
\newcommand{\om}{\omega}
\newcommand{\Om}{\Omega}

\newcommand{\ad}{{a^\dagger}}
\newcommand{\ada}{a^\dagger a}
\newcommand{\D}{\Delta}

\newcommand{\Hqd}{H_{\text{qd}}}

\newcommand{\Sig}{\Sigma_{21}}
\newcommand{\qdq}{{q}^\dagger q}
\newcommand{\qd}{{q^\dagger}}
\newcommand{\hc}{\mathrm{h.c.}}
\newcommand\mr{\mathrm}
\newcommand{\var}[1]{\ensuremath{\text{Var}[#1]}}

\newcommand{\avg}[1]{\left\langle #1 \right\rangle}		% average
\newcommand{\ie}{\textit{i.e.} }

\begin{document}

\title{Quantum squeezing in a nonlinear mechanical oscillator}

\author{Stefano Marti}
\thanks{These two authors contributed equally}
\author{Uwe von L\"upke}
\thanks{These two authors contributed equally}
\author{Om Joshi}
\author{Yu Yang}
\author{Marius Bild}
\author{Andraz Omahen}
\author{Yiwen Chu}
\author{Matteo Fadel}
\email[Corresponding author: ]{fadelm@phys.ethz.ch}
\affiliation{Department of Physics, ETH Z\"{u}rich, 8093 Z\"{u}rich, Switzerland}
\affiliation{Quantum Center, ETH Z\"{u}rich, 8093 Z\"{u}rich, Switzerland}

\date{\today}

\begin{abstract}
Mechanical degrees of freedom are natural candidates for continuous-variable quantum information processing and bosonic quantum simulations. These applications, however, require the engineering of squeezing and nonlinearities in the quantum regime.
Here we demonstrate ground state squeezing of a gigahertz-frequency mechanical resonator coupled to a superconducting qubit. This is achieved by parametrically driving the qubit, which results in an effective two-phonon drive. In addition, we show that the resonator mode inherits a nonlinearity from the off-resonant coupling with the qubit, which can be tuned by controlling the detuning.
We thus realize a mechanical squeezed Kerr oscillator, where we demonstrate the preparation of non-Gaussian quantum states of motion with Wigner function negativities and high quantum Fisher information.
This shows that our results also have applications in quantum metrology and sensing.
\end{abstract}

\maketitle

%\linenumbers

From the oscillation of a trapped particle to the vibration of a solid-state structure, mechanical modes are ubiquitous degrees of freedom that can exhibit sought-after properties such as high quality factors and large coupling rates to spins and electromagnetic fields. 
When operated in the quantum regime, mechanical modes are powerful building blocks for quantum technologies, with applications in information processing \cite{chamberland2022building,Qiao23}, bosonic simulations \cite{Chen2023}, memories \cite{Hann2019} and microwave-to-optical frequency conversion \cite{Mirhosseini2020,Delaney2022}. 
Moreover, their nonzero mass makes them particularly suited for sensing forces \cite{Biercuk10,Schreppler14,Ivanov16}, as well as for fundamental physics investigations, ranging from tests of the superposition principle \cite{macroPRL} to the detection of dark matter \cite{Carney21} and quantum gravity effects \cite{BonaldiQGtest}. 
To fully unlock these applications, however, it is necessary to have available a sophisticated toolbox for the preparation and manipulation of quantum states of motion, which is a nontrivial task.

In this context, mechanical resonators have recently attracted a lot of attention as new elements for hybrid quantum systems \cite{ClerkNatReview,Chu20}. 
In particular, gigahertz-frequency resonators can be interfaced to superconducting qubits and thus controlled with the toolbox of circuit quantum acoustodynamics (cQAD).
For example, resonant interaction with a qubit was used to demonstrate the preparation of mechanical Fock states \cite{Satzinger18,Arrangoiz-Arriola2018,Chu2018} and Schr\"odinger cat states \cite{catSCI23}. 
Crucially, compared to their electromagnetic counterparts, mechanical resonators have small physical footprints and a high density of accessible long-lived modes, making them ideal candidates for hardware-efficient quantum processors \cite{chamberland2022building} and quantum random access memories \cite{Hann2019}.

The realization of continuous variable (CV) quantum computing and bosonic simulations relies on the availability of a universal gate set composed of phase shift, displacement, beam-splitter, single-mode squeezing and Kerr nonlinearity \cite{LloydCVPRL19,MariPRL12,BraunsteinRMP,GaussQinfoRMP12}.
The first two are relatively simple to realize through free evolution and coherent driving. Beam-splitter operations have been recently demonstrated in cQAD between surface \cite{Qiao23} and bulk \cite{BSpaper} acoustic waves.
For mechanical systems, quantum noise squeezing was pioneered in trapped ions \cite{MeekhofPRL96}, and later demonstrated in drum oscillators using the tools of electromechanics \cite{WollmanSCI15,PirkkalainenPRL15,LecocqPRX15,DelaneyPRL19,youssefi2023squeezed}.
In cQAD, a recent experiment demonstrated two-mode squeezing of gigahertz-frequency surface acoustic waves through modulation of one of the Bragg reflectors \cite{SAWsqPRXQ22}. 
Nonlinear evolutions in the quantum regime are difficult to realise with standard opto- or electromechanical coupling, since this coupling is linear for small displacements. 
One possibility is to off-resonantly couple a mechanical oscillator to a two-level system, which gives rise to an effective nonlinearity for the phonon through a hybridization of the modes. 
Recently, this was demonstrated in an experiment where the vibrational modes of a carbon nanotube were coupled to a quantum dot \cite{SamantaNatPhys}.
Despite all this progress, however, the demonstration of a full gate set for universal CV quantum information processing in a single cQAD device is still lacking.

\begin{figure*}[tb]
\centering
\includegraphics[width=\textwidth]{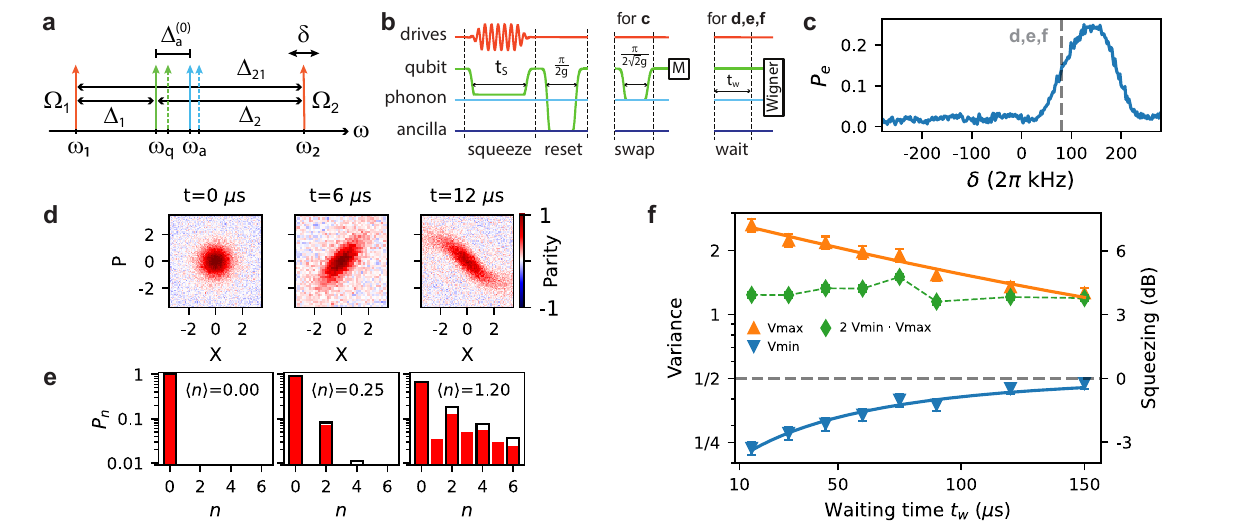}
\caption{\textbf{Preparation of squeezed states in a HBAR.}
\textbf{(a)} Schematic illustration of the spectrum used for parametric squeezing. The application of the two parametric drives (orange) results in an AC Stark shift of the qubit, which in turn changes the phonon frequency (dashed lines), therefore requiring a correction $\delta$ to $\omega_2$ in order to meet the squeezing condition $\omega_1+\omega_2=2\omega_a$.
\textbf{(b)} Pulse sequences used in the experiments. After applying the parametric drives to the system for time $t_S$, we reset the qubit and implement either a measurement of the phonon mode population (for panel c), or a measurement of the Wigner function after a wait time $t_\mr{w}$ (for panels d,e,f). The time $t_\mr{w}$ is zero, except for the measurement in panel f. 
\textbf{(c)} Measurement of the qubit population after a swap operation with the phonon mode, for different $\delta$. The peak signals when excitations are created in the mechanical mode. The gray dashed line at $\delta=2\pi\cdot\unit{80}{kHz}$ indicates the setting used for the measurements in panels d, e and f.
\textbf{(d)} Wigner functions of the phonon mode for $t_S=\unit{0, 6, 12}{\mu s}$, respectively. 
\textbf{(e)} Fock state populations of the states shown in panel d, extracted from maximum likelihood reconstructions.
\textbf{(f)} Decay of a squeezed state observed from the evolution of $V_\text{min}$ and $V_\text{max}$ measured for a variable wait time $t_\mr{w}$ after state preparation. Continuous lines are fits to the data (see main text), while the dashed horizontal line indicates the ground state variance.
}
\label{fig1}
\end{figure*}

In this work, we present ground state squeezing of a gigahertz-frequency phonon mode of a high-overtone bulk acoustic wave resonator (HBAR) with tuneable nonlinearity. 
The phonon mode is coupled to a superconducting qubit, which we use as a mixing element for implementing the effective squeezing drive: by applying two microwave tones to the qubit, we activate a parametric process that creates pairs of phonons in the resonator. Moreover, this coupling gives rise to an effective Kerr nonlinearity for the phonon mode, which we tune by changing the phonon-qubit detuning.
To characterize our system, we study the dependence of the squeezing rate and of the Kerr nonlinearity on different system parameters. Having demonstrated control over both these quantities, we combine them to realize a mechanical version of a squeezed Kerr oscillator, a paradigmatic model in quantum optics. 
By using the qubit to perform direct Wigner function measurements, we show that operating this system in different regimes results in the preparation of non-Gaussian states of motion with Wigner negativities and high quantum Fisher information.

\vspace{2mm}
The device used in this work is a cQAD system where a transmon qubit is flip-chip bonded to a HBAR, an improved version of devices used in previous works \cite{Chu2018, vonLupke22}.
The qubit has a frequency $\omega_q = 2\pi\cdot\unit{5.042}{GHz}$, which can be tuned via a Stark shift drive \cite{vonLupke22}. 
At this frequency, the qubit has an energy relaxation time $T_1=\unit{17(0.4)}{\mu s}$, Ramsey decoherence time $T_2^\ast=\unit{24(0.7)}{\mu s}$, and anharmonicity $\alpha=2\pi\cdot\unit{185}{MHz}$. 
The HBAR is coupled to the qubit through a piezoelectric transducer made of aluminum nitride that mediates a Jaynes-Cummings (JC) interaction with a coupling strength $g=2\pi\cdot\unit{292}{kHz}$. 
The phonon mode we consider in this work has a frequency $\omega_a=2\pi\cdot\unit{5.023}{GHz}$, an energy relaxation time $T_1=\unit{132(4)}{\mu s}$ and a Ramsey decoherence time $T_2^\ast=\unit{210(9)}{\mu s}$.

\vspace{2mm}

Our system can be described by the Hamiltonian 

\begin{align}
    H_{\text{cQAD}}/\hbar &= \omega_q \qd q - \dfrac{\alpha}{2} {\qd}^2 q^2 \notag\\
    &\quad + \omega_a \ad a + g(q \ad + \qd a) + \Hqd/\hbar \;, \label{eq:fullH}
\end{align}
where $q$ and $a$ are the bosonic annihilation operators for the qubit and the phonon mode, respectively. 
The term $H_{\text{qd}}/\hbar=(\Omega_1 e^{-i\omega_1 t} + \Omega_2 e^{-i\omega_2 t}) q^\dagger+\text{h.c.}$ describes the two off-resonant microwave drives at frequencies $\omega_{1,2}$ and amplitude $\Omega_{1,2}$ applied to the qubit, see Fig.~\ref{fig1}a. 
We define the detunings $\Delta_{1,2}=\omega_{1,2}-\omega_q$ and the dimensionless drive strengths $\xi_{1,2}=\Omega_{1,2}/\Delta_{1,2}$. In addition, we use a third, far off-resonant drive to control the qubit frequency via the AC Stark shift. 

When the resonance condition $\omega_1+\omega_2 = 2\omega_a$ is fulfilled, the qubit nonlinearity mediates a four-wave mixing process that results in a two-phonon drive (${\ad}^2 + a^2$). The emergence of this squeezing term can be unveiled through a series of unitary transformations (see \cite{SM} for details). 
In combination with the nonlinearity the phonon mode inherits from the qubit, this results in an effective squeezed Kerr Hamiltonian for the phonon mode  
\begin{equation}\label{eq:HsqKerr}
    H/\hbar = -\Delta \ad a - \epsilon ({\ad}^2 + a^2) - K {\ad}^2 a^2 \;.
\end{equation}
Here, $\Delta= ( \omega_1 + \omega_2 - 2\omega_a' )/2$,  where $\omega_a' \approx \omega_a + \frac{g^2}{\D_a}$, is the frequency of the phonon mode including a normal mode shift due to the presence of the qubit. $\D_a=\omega_a - \omega_q^{ss}$ is the detuning between the phonon mode and the AC Stark shifted qubit.
The squeezing rate $\epsilon$ is given by \cite{zhang2019engineering,Wang2019,SM} 
\begin{equation}\label{eq:epsilon}
    \epsilon = 2 \dfrac{g^2}{\D_a}\xi_1\xi_2 \dfrac{\alpha}{\Sig + \alpha}\;,
\end{equation}
where $\Sig=\D_1+\D_2$. 
Finally, $K$ is the Kerr nonlinearity, which is $K\approx g^4/\Delta_a^3$ for $\alpha\gg\Delta_a \gg g$ \cite{SM}. 

The Hamiltonian in Eq.~\eqref{eq:HsqKerr} is a paradigmatic model in quantum optics, exhibiting a plethora of interesting phenomena such as chaotic dynamics \cite{MilburnPRA91}, quantum phase transitions \cite{CCarlosNPJ23}, tunneling \cite{WielingaPRA93}, and parametric amplification \cite{Boutin17}. 
Moreover, this model admits macroscopic superpositions as quantum ground states, which can be exploited for error-protected qubit encoding \cite{CochranePRA99,Goto16,Puri17}. 
The latter application made squeezed Kerr oscillators particularly attractive for quantum information processing, which motivated their recent experimental implementation for electromagnetic modes in circuit QED platforms~\cite{ZakisSCI15,Grimm2020,frattini2022squeezed,Iyama23}. 
Here we present the implementation for a mechanical mode, and use it to prepare squeezed and non-Gaussian quantum states of motion of a massive system.

\begin{figure}[tb]
\centering
\includegraphics[width=8cm]{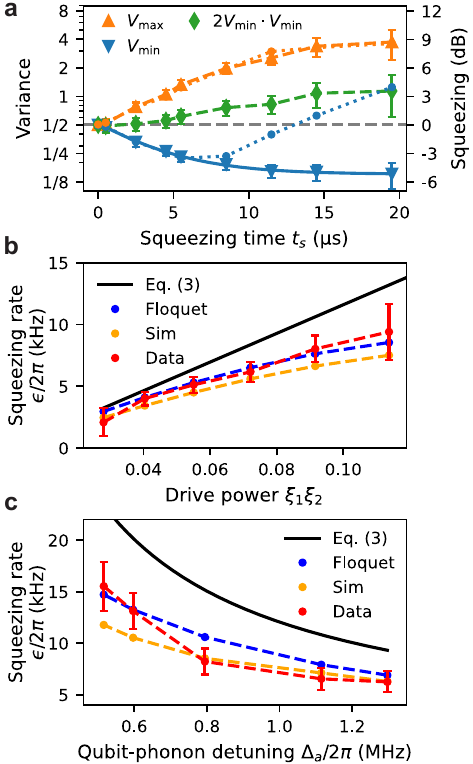}
\caption{\textbf{Squeezing rate characterization.}
\textbf{(a)} Example of a measured evolution of $V_\text{min}$ and $V_\text{max}$, see main text for the experimental parameters. Dashed and dotted lines are guides to the eye for the variances extracted by Gaussian fit (triangles) and state reconstruction (dots), respectively. These two methods agree for short times $t_S\lesssim\unit{7}{\mu s}$, where the state is close to Gaussian.
We extract the squeezing rate $\epsilon$ by fitting the evolution of $V_\text{min}$ with a decaying exponential model (see main text), indicated by the blue solid line.
\textbf{(b)} Measured dependence of the squeezing rate as a function of the drive powers, for $\Delta_a=\unit{2\pi\cdot 1.5}{MHz}$. The black line is the prediction obtained from Eq.~\eqref{eq:epsilon}. The yellow points are obtained from a time-dependent simulation of the squeezing dynamics. The blue points are obtained from a Floquet simulation \cite{zhang2019engineering}. Dashed lines are guides to the eye.
\textbf{(c)} Measured dependence of the squeezing rate as a function of the qubit-phonon detuning $\Delta_a$, for $\xi_1\xi_2=0.07$. Points and lines are as in panel b.
In all panels, $\xi_1\approx \xi_2$.}
\label{fig2}
\end{figure}

\vspace{2mm}
First, let us investigate the squeezing dynamics for a large $\Delta_a= 2\pi \cdot 1.5\,$MHz, such that $K\propto \Delta_a^{-3}$ is significantly smaller than $\epsilon\propto\Delta_a^{-1}$. 
When the parametric drives are applied, the qubit frequency is modified by the AC Stark shift, which then results in a change to the normal mode shift of the phonon mode, see Fig.~\ref{fig1}a. 
To ensure that the resonance condition for squeezing is satisfied, we perform the following calibration experiment, see Fig.~\ref{fig1}b: 
We apply the parametric drives for a time $t_S=\unit{20}{\mu s}$, including \unit{0.5}{\mu s} Gaussian edges, with drive frequency $\omega_2=2\omega_a-\omega_1+\delta$.
We then reset the qubit to its ground state by swapping the population acquired during the off-resonant driving to an ancillary phonon mode. 
Finally, we bring the qubit into resonance with the phonon mode we want to squeeze for a time $\pi/(2\sqrt{2}g)$, thereby swapping part of the phonon population to the qubit, after which we measure the qubit state using standard dispersive readout. 
Repeating this experiment for different $\delta$ results in Fig.~\ref{fig1}c, showing a peak around $\delta\approx2\pi\cdot\unit{140}{kHz}$ that provides an indication of when the two-phonon drive becomes resonant.

To fully characterize the state resulting from the two-phonon drive and verify the coherence of this process, we set a desired $\delta$ and $t_S$, and perform a Wigner function measurement of the phonon mode \cite{vonLupke22}. 
The results for $\delta=2\pi\cdot\unit{80}{kHz}$ and $t_S=\unit{0, 6, 12}{\mu s}$ are shown in Fig.~\ref{fig1}d. 
For $t_S=0\,\mu$s we obtain a measurement of the ground state, while for larger times we observe a reduction of the quantum noise along one quadrature, \ie squeezing, as well as an increase in the perpendicular quadrature. 
Note that for the longest evolution time we see also a distortion of the state, that is due to the residual phonon mode nonlinearity. As we will see later in more detail, this distortion limits the squeezing, and it depends on $\delta$. Choosing $\delta=2\pi\cdot\unit{80}{kHz}$ allowed us to observe the strongest squeezing, as a result of a partial compensation of the nonlinearity $K$ by the detuning $\Delta\approx g^2/\Delta_a - \delta/2$ in Eq.~\eqref{eq:HsqKerr}. 

Given phase space quadratures $X$ and $P$, we define the variance along direction $\theta$ as $V(\theta)=\text{Var}[X \cos\theta + P \sin\theta]$. 
For an ideal ground state $V_\text{GS}=V(\theta)=1/2$, therefore observing a smaller value implies quantum squeezing. 
We thus define $V_\text{min}=\min_\theta V(\theta)$ and $V_\text{max}=\max_\theta V(\theta)$ as the squeezed and antisqueezed quadratures.
Fitting the Wigner function at $t_S=\unit{6}{\mu s}$ with a two-dimensional Gaussian, we obtain $V_\text{min}=0.252(6)$, which corresponds to a noise reduction of $\unit{3.0(1)}{dB}$ below $V_\text{GS}$. From this, and $V_\text{max}=1.45(4)$, we estimate a thermal population of $\sqrt{V_\text{min}V_\text{max}}-1/2=0.10(1)$, corresponding to a state purity of $83(1)\%$ \cite{SM}. For the chosen experimental parameters, numerical simulations indicate that $t_S\approx\unit{6}{\mu s}$ is the optimal time for the strongest squeezing, this being mostly limited by residual nonlinearity (see later Fig.~\ref{fig2}a). In fact, longer evolution times result in non-Gaussian states, such as the one shown for $t_S=\unit{12}{\mu s}$, for which the above analysis is not reliable.

Alternatively, maximum-likelihood estimate allows us to reconstruct from the measurements shown in Fig.~\ref{fig1}d the corresponding density matrices \cite{Chou2018}, from which we calculate the associated covariance matrices. For $t_S=\unit{6}{\mu s}$ we obtain $V_\text{min}=0.236(1)$, while for $t_S=\unit{12}{\mu s}$ we obtain $V_\text{min}=0.268(3)$, which shows less squeezing due to the nonlinear evolution.
We then plot the diagonal elements of the density matrices in Fig.~\ref{fig1}e, corresponding to the Fock state populations of the phonon states. 
We observe high populations of even states, characteristic of a two-phonon drive. 
Black bars indicate a fit of the reconstructed populations to the closest ideal squeezed state.

\begin{figure}[tb]
\centering
\includegraphics[width=0.8\columnwidth]{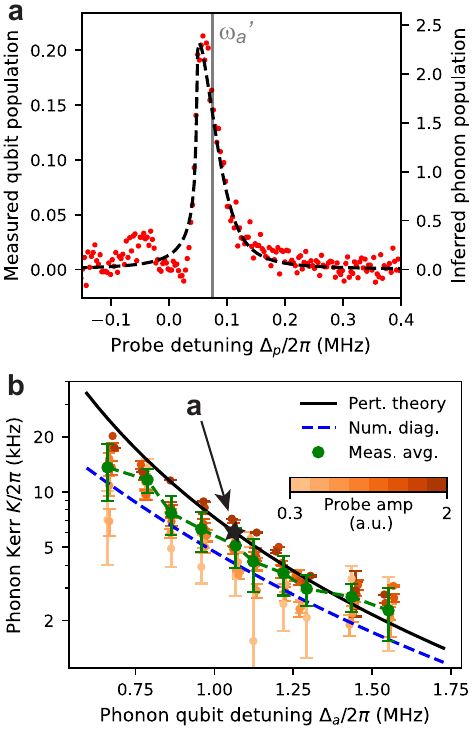}
\caption{\textbf{Phonon mode anharmonicity characterization.}
\textbf{(a)} Phonon mode spectroscopy showing an asymmetric peak, characteristic of a nonlinear oscillator. Red dots show the measured qubit population when probing at different frequencies. From the measured qubit population we infer the steady state population of the phonon mode, which we fit to the solution of a classical driven Duffing oscillator to extract the mode anharmonicity (see main text and \cite{SM} for details). The vertical gray line marks the fitted frequency of the oscillator. 
\textbf{(b)} Phonon nonlinearity extracted using the method shown in panel a for different qubit-phonon detunings $\Delta_a$ and probe amplitudes (orange points), compared to the results obtained from exact diagonalization of Eq.~\eqref{eq:fullH} without drives (blue dashed line) and to the analytical expression obtained from perturbation theory $g^4/\Delta_a^3$ (black line, see \cite{SM} for details).
}
\label{fig3}
\end{figure}

To characterize the usefulness of the prepared squeezed states for applications such as quantum metrology, we measure their lifetime. For this, we prepare a squeezed state and wait for a variable time $t_\mr{w}$ before performing the Wigner function measurement, see Fig.~\ref{fig1}b. 
The (anti)squeezing for different $t_\mr{w}$ is shown in Fig.~\ref{fig1}f. As expected from the free evolution of a squeezed state in the presence of relaxation, we observe that its variances gradually return to those of the ground state \cite{SM}. 
Fitting the squeezed variance measurements with $(1-e^{-\gamma_d t}(1-2 V_\text{min}))/2$, where $V_\text{min}$ is the variance at $t_\mr{w}=0$, gives a decay time $\gamma_d^{-1}=\unit{78(11)}{\mu s}$.
While the decay time of $V_\text{max}$ of $\unit{125(12)}{\mu s}$ is compatible with the phonon $T_1$, we attribute the lower $\gamma_d^{-1}$ to the fact that the squeezed quadrature is more sensitive to dephasing than the antisqueezed quadrature.

We now investigate the dependence of the squeezing rate $\epsilon$ on the experimental parameters. We extract $\epsilon$ by measuring the evolution of $V_\text{min}$ as a function of the squeezing time $t_S$.
Concretely, $\epsilon$ is inferred from a fit of the squeezing measurements with the function $V_\text{min}(t)=(\gamma +4 \epsilon  e^{-t (\gamma +4 \epsilon )})/2(\gamma + 4 \epsilon )$, which describes the squeezing dynamics in the presence of decay at rate $\gamma$ \cite{SM}. 
Note that $\epsilon$ can be estimated from the evolution at short times $t_S\ll 1/K$, where the state is still Gaussian. For this reason we obtain $V_\text{min}$ from a Gaussian fit, after having checked that for short times it is consistent with the one obtained from state reconstruction.
An example is shown in Fig.~\ref{fig2}a, for $\delta=2\pi\cdot\unit{80}{kHz}$, $\Delta_a=2\pi\cdot\unit{1.5}{MHz}$, $\xi_1=0.28$, and $\xi_2=0.26$. Fitting of $V_\text{min}(t)$ gives us an effective decay time $\gamma^{-1}=\unit{12.8(11)}{\mu s}$ and squeezing rate $\epsilon=2\pi\cdot\,\unit{7.6(3)}{kHz}$. Note that this effective decay time is shorter than the bare phonon $T_1$, as these measurements are subject to Purcell decay via the qubit and to additional dephasing resulting from finite qubit population and parametric driving.

\begin{figure*}[tb]
\centering
\includegraphics[width=\textwidth]{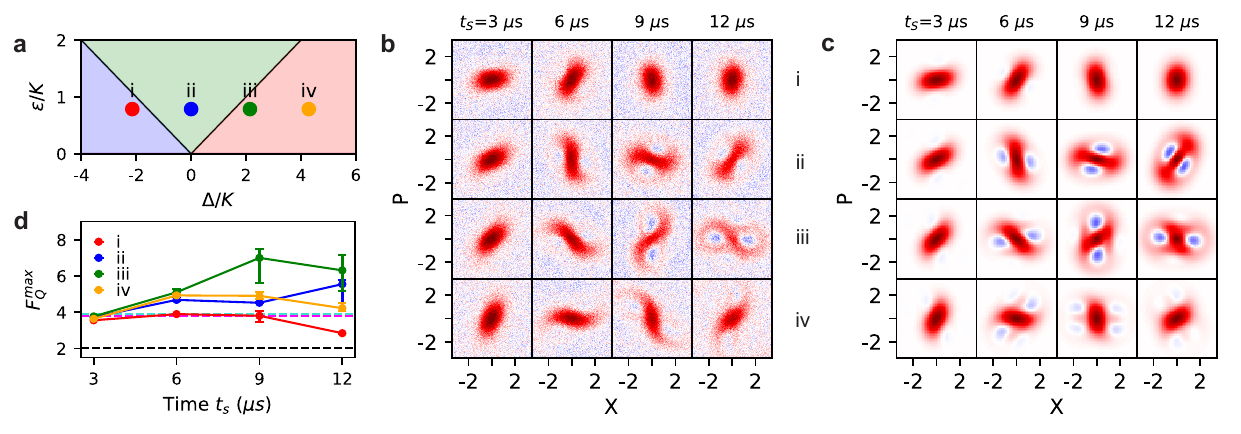}
\caption{\textbf{Preparation of non-Gaussian states of motion.}
\textbf{(a)} Parameter space of a squeezed Kerr oscillator. The three shaded regions correspond  to an effective single-, double- and triple-node potential (from left to right) in the semiclassical picture. The four points indicate the parameters used for the measurements in panel b.
\textbf{(b)} Wigner functions measured at the points shown in panel a, for different squeezing times $t_S$.
\textbf{(c)} Wigner functions obtained by simulating the effective Hamiltonian Eq.~\eqref{eq:HsqKerr} with $\epsilon = 2\pi\cdot 11 \,$kHz, $K=2\pi\cdot 14\,$kHz and including a decay rate of $(\unit{40}{\mu s})^{-1}$. 
\textbf{(d)} Quantum Fisher information for the states shown in panel b, computed from maximum likelihood reconstruction of the density matrix. The black dashed line indicates the theoretical value achieved by coherent states, representing the classical limit. The turquoise and magenta dashed lines indicate the experimental values achieved by a Fock $\ket{1}$ state and by the largest cat state of Ref.~\cite{catSCI23}, respectively.
}
\label{fig4}
\end{figure*}

Repeating the measurement shown in Fig.~\ref{fig2}a for different drive powers $\xi_1 \xi_2$ and qubit-phonon detunings $\Delta_a$ we obtain the rates shown in Fig.~\ref{fig2}b, c. The precise value of these controllable parameters is extracted from independent measurements in the following way. 
We set the desired drive strengths $\xi_{1,2}$, which we previously calibrated via their Stark shift on the qubit \cite{BSpaper}. 
The detuning $\Delta_a$ is then set by changing the qubit frequency via the independent Stark shift drive. 
In Fig.~\ref{fig2}b, c we compare the measured squeezing rate $\epsilon$, to the one predicted by Eq.~\eqref{eq:epsilon}. 
We see a disagreement that we attribute to effects of higher order in $\xi_1\xi_2$, which are not included in Eq.~\eqref{eq:epsilon}. 
For this reason, we add a comparison with the rates expected from Floquet theory \cite{zhang2019engineering} and from a time-domain simulation of our system Hamiltonian, Eq. \eqref{eq:fullH}, which shows good agreement with the measurements.

After having characterized the squeezing rate, we proceed with investigating the nonlinearity $K$ of the phonon mode. 
In Fig.~\ref{fig3}a we show a spectroscopic measurement taken by applying a \unit{400}{\mu s} long probe tone detuned by $\D_p = \omega_p - \omega_a$ from the phonon mode and then measuring the qubit state population.
When the probe tone is resonant with the phonon mode, it drives the phonon mode into a steady state, from which population leaks to the qubit, which results in our spectroscopic readout signal. 
This lets us infer the population of the phonon mode \cite{SM}, which we denote on the second y-axes of Fig.~\ref{fig3}a. 
We observe an asymmetric resonance peak characteristic of a nonlinear Duffing oscillator. 
Fitting these data with the equation of motion of a classical driven Duffing oscillator \cite{LifshitzBook}, we extract the nonlinearity of the mode \cite{SM}.  
Repeating this analysis for different probe amplitudes and detunings $\Delta_a$ gives us the data shown in Fig.~\ref{fig3}b. 
These are in good agreement with the predictions obtained from numerical diagonalization of Eq.~\eqref{eq:fullH} without drives, as well as with the analytical expression $K\approx g^4/\Delta_a^3$ obtained from fourth-order perturbation theory \cite{SM}. This shows that we can tune the phonon nonlinearity by approximately one order of magnitude.

Having demonstrated control over the parameters of Eq.~\eqref{eq:HsqKerr}, we now show that different regimes of this Hamiltonian can be used to prepare non-Gaussian states. 
The dynamics of the system is fully determined by the two dimensionless parameters $\Delta/K$ and $\epsilon/K$, where the former is controlled by the drive frequencies and the latter by the drive amplitudes. 
This two-dimensional parameter space is divided into three different regions by two phase transitions at $\Delta=\pm 2 \epsilon$ \cite{DykmanPRE98,WustmanPRB13,Venkatraman}. 
In a semiclassical description, these regions are associated with an effective single-, double-, and triple-node potential in the frame rotating at half the driving frequency \cite{Venkatraman,EichlerBook23}, see Fig.~\ref{fig4}a.

To prepare interesting mechanical states, we start with the phonon mode in the ground state, and then let it evolve according to Hamiltonian Eq.~\eqref{eq:HsqKerr}. 
We choose $\xi_{1}\xi_2=0.07$ and $\D_a=2\pi\cdot \unit{0.53}{MHz}$.
This fixes $\epsilon/K$, but leaves us control over $\Delta/K$ by changing the parametric drive correction $\delta$. 
The Wigner functions measured at different values of $t_S$ (see Fig. \ref{fig1}b) are shown in Fig. \ref{fig4}b. 
We note parameter regimes that result in states significantly deviating from Gaussian states. 
Moreover, in some cases negative Wigner function regions appear, which are a direct indication of non-classicality \cite{KenfackJOB2004,MattiaPRXQ}. The states we observe can be qualitatively understood from an evolution of the ground state in the semiclassical potential associated with the chosen parameter regime. From exact diagonalization we obtain $K=2\pi\cdot 14(1)\,$kHz, while from a Gaussian fit of the states at $t_S=\unit{3}{\mu s}$ we estimate $\epsilon = 2\pi\cdot 11(1)\,$kHz. 
We show in Fig.~\ref{fig4}c numerical simulations of Eq.~\eqref{eq:HsqKerr} with these parameters, which show good agreement with our measurements. For these simulations, we use a lower phonon lifetime of $\unit{40}{\mu s}$, estimated taking into account Purcell decay via the qubit, and include a global rotation to take into account that the measurements in Fig.~\ref{fig4}b are performed in the rotating frame of the qubit.

To have a pragmatic characterization of the states we can prepare, we quantify their usefulness for a metrological protocol by means of the quantum Fisher information (QFI). 
For a perturbation generated by the operator $\hat{A}$, the QFI associated with the state $\rho$ is defined as $F_Q[\rho,A] = 2 \sum_{k,l}  \frac{(\lambda_k-\lambda_l)^2}{(\lambda_k+\lambda_l)}\vert\bra{k}A\ket{l}\vert^2$, where $\lambda_k$ and $\ket{k}$ are the eigenvalues and eigenvectors of $\rho$, respectively, and the summation goes over all $k,l$ such that $\lambda_k+\lambda_l>0$. 
Taking the parameter estimation task to be the measurement of a displacement amplitude, then $A(\theta) = X \sin\theta + P \cos\theta$, with $\theta$ specifying the displacement direction. From this we define $F_Q^\text{max}=\max_\theta F_Q[\rho,A(\theta)]$, which gives the maximum sensitivity attainable by the state. 
Note that a coherent state has $F_Q[\ket{\alpha},A(\theta)]=2$, meaning that if we consider coherent states as classical resources then any $F_Q^\text{max}>2$ implies non-classicality.
We estimate $F_Q^\text{max}$ numerically by first reconstructing the density matrix of the state, and then maximizing $F_Q[\rho,A(\theta)]$ over $\theta$. 
The results are shown in Fig.~\ref{fig4}d, together with the values we obtain for a Fock $\ket{1}$ state and for the Schr\"{o}dinger cat states of Ref.~\cite{catSCI23}. 
The Fisher information is computed for the time-evolution of states at four different detunings. 
We observe that the states corresponding to point $iii$ in Fig.~\ref{fig4}a exhibit significantly larger values after \unit{6}{\mu s} than previously measured states, while the states in the squeezed regime do not surpass them.\\

% \vspace{2mm}
In conclusion, we have demonstrated ground state squeezing of a gigahertz-frequency phonon mode of a HBAR device with tunable nonlinearity. This allows us to prepare non-Gaussian states of motion characterized by a high quantum Fisher information, which can find immediate application in quantum sensing with mechanical degrees of freedom. Our results, in combination with the beam-splitter operation demonstrated in the same platform in Ref.~\cite{BSpaper}, complete the toolbox for universal CV quantum information processing and bosonic quantum simulation in HBARs. This opens up the possibility to use the large number of modes available in these devices for hardware-efficient quantum chemistry simulations \cite{Wang2019,Chen2023}, as well as for nonlinear boson sampling \cite{nlBSamp2023}.\\

% \vspace{2mm}
{\centering\textbf{Acknowledgements}\\}
\noindent The authors thank Alex Eichler, Francesco Adinolfi and Hugo Doeleman for useful discussions and feedback on the manuscript, and Max Drimmer for contributing to the device fabrication. Fabrication of the device was performed at the FIRST cleanroom of ETH Z\"urich and the BRNC cleanroom of IBM Z\"urich. We acknowledge support from the Swiss National Science Foundation under grant $200021\_204073$. MF was supported by the Swiss National Science Foundation Ambizione Grant No. 208886, and The Branco Weiss Fellowship -- Society in Science, administered by the ETH Z\"{u}rich.

\vspace{2mm}
{\centering\textbf{Author contributions}\\}
\noindent UvL and MF conceived the experiments. UvL, YY, MB, and AO fabricated the device. SM, UvL, YY, and MF wrote experiment control sequences. SM, UvL, and MF performed measurements and analysed the data. SM, UvL, OJ, and MF performed numerical simulations of the experiments. UvL and MF derived theoretical models. YC and MF supervised the work. SM, UvL, and MF wrote the manuscript with feedback from all authors.

\vspace{2mm}
{\centering\textbf{Data and code availability}\\}
\noindent Raw data and analysis scripts will be made available on Zenodo. Additional material is available from the corresponding author on reasonable request.\\

\bibliographystyle{apsrev4-1} 

\let\oldaddcontentsline\addcontentsline% Store \addcontentsline
\renewcommand{\addcontentsline}[3]{}% Make 
\bibliography{mybib}

\begin{thebibliography}{58}%
\makeatletter
\providecommand \@ifxundefined [1]{%
 \@ifx{#1\undefined}
}%
\providecommand \@ifnum [1]{%
 \ifnum #1\expandafter \@firstoftwo
 \else \expandafter \@secondoftwo
 \fi
}%
\providecommand \@ifx [1]{%
 \ifx #1\expandafter \@firstoftwo
 \else \expandafter \@secondoftwo
 \fi
}%
\providecommand \natexlab [1]{#1}%
\providecommand \enquote  [1]{``#1''}%
\providecommand \bibnamefont  [1]{#1}%
\providecommand \bibfnamefont [1]{#1}%
\providecommand \citenamefont [1]{#1}%
\providecommand \href@noop [0]{\@secondoftwo}%
\providecommand \href [0]{\begingroup \@sanitize@url \@href}%
\providecommand \@href[1]{\@@startlink{#1}\@@href}%
\providecommand \@@href[1]{\endgroup#1\@@endlink}%
\providecommand \@sanitize@url [0]{\catcode `\\12\catcode `\$12\catcode
  `\&12\catcode `\#12\catcode `\^12\catcode `\_12\catcode `\%12\relax}%
\providecommand \@@startlink[1]{}%
\providecommand \@@endlink[0]{}%
\providecommand \url  [0]{\begingroup\@sanitize@url \@url }%
\providecommand \@url [1]{\endgroup\@href {#1}{\urlprefix }}%
\providecommand \urlprefix  [0]{URL }%
\providecommand \Eprint [0]{\href }%
\providecommand \doibase [0]{http://dx.doi.org/}%
\providecommand \selectlanguage [0]{\@gobble}%
\providecommand \bibinfo  [0]{\@secondoftwo}%
\providecommand \bibfield  [0]{\@secondoftwo}%
\providecommand \translation [1]{[#1]}%
\providecommand \BibitemOpen [0]{}%
\providecommand \bibitemStop [0]{}%
\providecommand \bibitemNoStop [0]{.\EOS\space}%
\providecommand \EOS [0]{\spacefactor3000\relax}%
\providecommand \BibitemShut  [1]{\csname bibitem#1\endcsname}%
\let\auto@bib@innerbib\@empty
%</preamble>
\bibitem [{\citenamefont {Chamberland}\ \emph {et~al.}(2022)\citenamefont
  {Chamberland}, \citenamefont {Noh}, \citenamefont {Arrangoiz-Arriola},
  \citenamefont {Campbell}, \citenamefont {Hann}, \citenamefont {Iverson},
  \citenamefont {Putterman}, \citenamefont {Bohdanowicz}, \citenamefont
  {Flammia}, \citenamefont {Keller} \emph {et~al.}}]{chamberland2022building}%
  \BibitemOpen
  \bibfield  {author} {\bibinfo {author} {\bibfnamefont {C.}~\bibnamefont
  {Chamberland}}, \bibinfo {author} {\bibfnamefont {K.}~\bibnamefont {Noh}},
  \bibinfo {author} {\bibfnamefont {P.}~\bibnamefont {Arrangoiz-Arriola}},
  \bibinfo {author} {\bibfnamefont {E.~T.}\ \bibnamefont {Campbell}}, \bibinfo
  {author} {\bibfnamefont {C.~T.}\ \bibnamefont {Hann}}, \bibinfo {author}
  {\bibfnamefont {J.}~\bibnamefont {Iverson}}, \bibinfo {author} {\bibfnamefont
  {H.}~\bibnamefont {Putterman}}, \bibinfo {author} {\bibfnamefont {T.~C.}\
  \bibnamefont {Bohdanowicz}}, \bibinfo {author} {\bibfnamefont {S.~T.}\
  \bibnamefont {Flammia}}, \bibinfo {author} {\bibfnamefont {A.}~\bibnamefont
  {Keller}},  \emph {et~al.},\ }\href {\doibase
  http://dx.doi.org/10.1103/PRXQuantum.3.010329} {\bibfield  {journal}
  {\bibinfo  {journal} {PRX Quantum}\ }\textbf {\bibinfo {volume} {3}},\
  \bibinfo {pages} {010329} (\bibinfo {year} {2022})}\BibitemShut {NoStop}%
\bibitem [{\citenamefont {Qiao}\ \emph {et~al.}(2023)\citenamefont {Qiao},
  \citenamefont {Dumur}, \citenamefont {Andersson}, \citenamefont {Yan},
  \citenamefont {Chou}, \citenamefont {Grebel}, \citenamefont {Conner},
  \citenamefont {Joshi}, \citenamefont {Miller}, \citenamefont {Povey},
  \citenamefont {Wu},\ and\ \citenamefont {Cleland}}]{Qiao23}%
  \BibitemOpen
  \bibfield  {author} {\bibinfo {author} {\bibfnamefont {H.}~\bibnamefont
  {Qiao}}, \bibinfo {author} {\bibfnamefont {E.}~\bibnamefont {Dumur}},
  \bibinfo {author} {\bibfnamefont {G.}~\bibnamefont {Andersson}}, \bibinfo
  {author} {\bibfnamefont {H.}~\bibnamefont {Yan}}, \bibinfo {author}
  {\bibfnamefont {M.-H.}\ \bibnamefont {Chou}}, \bibinfo {author}
  {\bibfnamefont {J.}~\bibnamefont {Grebel}}, \bibinfo {author} {\bibfnamefont
  {C.~R.}\ \bibnamefont {Conner}}, \bibinfo {author} {\bibfnamefont {Y.~J.}\
  \bibnamefont {Joshi}}, \bibinfo {author} {\bibfnamefont {J.~M.}\ \bibnamefont
  {Miller}}, \bibinfo {author} {\bibfnamefont {R.~G.}\ \bibnamefont {Povey}},
  \bibinfo {author} {\bibfnamefont {X.}~\bibnamefont {Wu}}, \ and\ \bibinfo
  {author} {\bibfnamefont {A.~N.}\ \bibnamefont {Cleland}},\ }\href {\doibase
  10.1126/science.adg8715} {\bibfield  {journal} {\bibinfo  {journal}
  {Science}\ }\textbf {\bibinfo {volume} {380}},\ \bibinfo {pages} {1030}
  (\bibinfo {year} {2023})}\BibitemShut {NoStop}%
\bibitem [{\citenamefont {Chen}\ \emph {et~al.}(2023)\citenamefont {Chen},
  \citenamefont {Lu}, \citenamefont {Zhang}, \citenamefont {Zhang},
  \citenamefont {Huang}, \citenamefont {Qiao}, \citenamefont {Su},
  \citenamefont {Zhang}, \citenamefont {Zhang}, \citenamefont {Banchi},
  \citenamefont {Kim},\ and\ \citenamefont {Kim}}]{Chen2023}%
  \BibitemOpen
  \bibfield  {author} {\bibinfo {author} {\bibfnamefont {W.}~\bibnamefont
  {Chen}}, \bibinfo {author} {\bibfnamefont {Y.}~\bibnamefont {Lu}}, \bibinfo
  {author} {\bibfnamefont {S.}~\bibnamefont {Zhang}}, \bibinfo {author}
  {\bibfnamefont {K.}~\bibnamefont {Zhang}}, \bibinfo {author} {\bibfnamefont
  {G.}~\bibnamefont {Huang}}, \bibinfo {author} {\bibfnamefont
  {M.}~\bibnamefont {Qiao}}, \bibinfo {author} {\bibfnamefont {X.}~\bibnamefont
  {Su}}, \bibinfo {author} {\bibfnamefont {J.}~\bibnamefont {Zhang}}, \bibinfo
  {author} {\bibfnamefont {J.-N.}\ \bibnamefont {Zhang}}, \bibinfo {author}
  {\bibfnamefont {L.}~\bibnamefont {Banchi}}, \bibinfo {author} {\bibfnamefont
  {M.~S.}\ \bibnamefont {Kim}}, \ and\ \bibinfo {author} {\bibfnamefont
  {K.}~\bibnamefont {Kim}},\ }\href {\doibase 10.1038/s41567-023-01952-5}
  {\bibfield  {journal} {\bibinfo  {journal} {Nature Physics}\ }\textbf
  {\bibinfo {volume} {19}},\ \bibinfo {pages} {877} (\bibinfo {year}
  {2023})}\BibitemShut {NoStop}%
\bibitem [{\citenamefont {Hann}\ \emph {et~al.}(2019)\citenamefont {Hann},
  \citenamefont {Zou}, \citenamefont {Zhang}, \citenamefont {Chu},
  \citenamefont {Schoelkopf}, \citenamefont {Girvin},\ and\ \citenamefont
  {Jiang}}]{Hann2019}%
  \BibitemOpen
  \bibfield  {author} {\bibinfo {author} {\bibfnamefont {C.~T.}\ \bibnamefont
  {Hann}}, \bibinfo {author} {\bibfnamefont {C.-L.}\ \bibnamefont {Zou}},
  \bibinfo {author} {\bibfnamefont {Y.}~\bibnamefont {Zhang}}, \bibinfo
  {author} {\bibfnamefont {Y.}~\bibnamefont {Chu}}, \bibinfo {author}
  {\bibfnamefont {R.~J.}\ \bibnamefont {Schoelkopf}}, \bibinfo {author}
  {\bibfnamefont {S.~M.}\ \bibnamefont {Girvin}}, \ and\ \bibinfo {author}
  {\bibfnamefont {L.}~\bibnamefont {Jiang}},\ }\href {\doibase
  10.1103/PhysRevLett.123.250501} {\bibfield  {journal} {\bibinfo  {journal}
  {Phys. Rev. Lett.}\ }\textbf {\bibinfo {volume} {123}},\ \bibinfo {pages}
  {250501} (\bibinfo {year} {2019})}\BibitemShut {NoStop}%
\bibitem [{\citenamefont {Mirhosseini}\ \emph {et~al.}(2020)\citenamefont
  {Mirhosseini}, \citenamefont {Sipahigil}, \citenamefont {Kalaee},\ and\
  \citenamefont {Painter}}]{Mirhosseini2020}%
  \BibitemOpen
  \bibfield  {author} {\bibinfo {author} {\bibfnamefont {M.}~\bibnamefont
  {Mirhosseini}}, \bibinfo {author} {\bibfnamefont {A.}~\bibnamefont
  {Sipahigil}}, \bibinfo {author} {\bibfnamefont {M.}~\bibnamefont {Kalaee}}, \
  and\ \bibinfo {author} {\bibfnamefont {O.}~\bibnamefont {Painter}},\ }\href
  {\doibase 10.1038/s41586-020-3038-6} {\bibfield  {journal} {\bibinfo
  {journal} {Nature}\ }\textbf {\bibinfo {volume} {588}},\ \bibinfo {pages}
  {599} (\bibinfo {year} {2020})}\BibitemShut {NoStop}%
\bibitem [{\citenamefont {Delaney}\ \emph {et~al.}(2022)\citenamefont
  {Delaney}, \citenamefont {Urmey}, \citenamefont {Mittal}, \citenamefont
  {Brubaker}, \citenamefont {Kindem}, \citenamefont {Burns}, \citenamefont
  {Regal},\ and\ \citenamefont {Lehnert}}]{Delaney2022}%
  \BibitemOpen
  \bibfield  {author} {\bibinfo {author} {\bibfnamefont {R.~D.}\ \bibnamefont
  {Delaney}}, \bibinfo {author} {\bibfnamefont {M.~D.}\ \bibnamefont {Urmey}},
  \bibinfo {author} {\bibfnamefont {S.}~\bibnamefont {Mittal}}, \bibinfo
  {author} {\bibfnamefont {B.~M.}\ \bibnamefont {Brubaker}}, \bibinfo {author}
  {\bibfnamefont {J.~M.}\ \bibnamefont {Kindem}}, \bibinfo {author}
  {\bibfnamefont {P.~S.}\ \bibnamefont {Burns}}, \bibinfo {author}
  {\bibfnamefont {C.~A.}\ \bibnamefont {Regal}}, \ and\ \bibinfo {author}
  {\bibfnamefont {K.~W.}\ \bibnamefont {Lehnert}},\ }\href {\doibase
  10.1038/s41586-022-04720-2} {\bibfield  {journal} {\bibinfo  {journal}
  {Nature}\ }\textbf {\bibinfo {volume} {606}},\ \bibinfo {pages} {489}
  (\bibinfo {year} {2022})}\BibitemShut {NoStop}%
\bibitem [{\citenamefont {Biercuk}\ \emph {et~al.}(2010)\citenamefont
  {Biercuk}, \citenamefont {Uys}, \citenamefont {Britton}, \citenamefont
  {VanDevender},\ and\ \citenamefont {Bollinger}}]{Biercuk10}%
  \BibitemOpen
  \bibfield  {author} {\bibinfo {author} {\bibfnamefont {M.~J.}\ \bibnamefont
  {Biercuk}}, \bibinfo {author} {\bibfnamefont {H.}~\bibnamefont {Uys}},
  \bibinfo {author} {\bibfnamefont {J.~W.}\ \bibnamefont {Britton}}, \bibinfo
  {author} {\bibfnamefont {A.~P.}\ \bibnamefont {VanDevender}}, \ and\ \bibinfo
  {author} {\bibfnamefont {J.~J.}\ \bibnamefont {Bollinger}},\ }\href {\doibase
  10.1038/nnano.2010.165} {\bibfield  {journal} {\bibinfo  {journal} {Nature
  Nanotechnology}\ }\textbf {\bibinfo {volume} {5}},\ \bibinfo {pages} {646}
  (\bibinfo {year} {2010})}\BibitemShut {NoStop}%
\bibitem [{\citenamefont {Schreppler}\ \emph {et~al.}(2014)\citenamefont
  {Schreppler}, \citenamefont {Spethmann}, \citenamefont {Brahms},
  \citenamefont {Botter}, \citenamefont {Barrios},\ and\ \citenamefont
  {Stamper-Kurn}}]{Schreppler14}%
  \BibitemOpen
  \bibfield  {author} {\bibinfo {author} {\bibfnamefont {S.}~\bibnamefont
  {Schreppler}}, \bibinfo {author} {\bibfnamefont {N.}~\bibnamefont
  {Spethmann}}, \bibinfo {author} {\bibfnamefont {N.}~\bibnamefont {Brahms}},
  \bibinfo {author} {\bibfnamefont {T.}~\bibnamefont {Botter}}, \bibinfo
  {author} {\bibfnamefont {M.}~\bibnamefont {Barrios}}, \ and\ \bibinfo
  {author} {\bibfnamefont {D.~M.}\ \bibnamefont {Stamper-Kurn}},\ }\href
  {\doibase 10.1126/science.1249850} {\bibfield  {journal} {\bibinfo  {journal}
  {Science}\ }\textbf {\bibinfo {volume} {344}},\ \bibinfo {pages} {1486}
  (\bibinfo {year} {2014})}\BibitemShut {NoStop}%
\bibitem [{\citenamefont {Ivanov}\ \emph {et~al.}(2016)\citenamefont {Ivanov},
  \citenamefont {Vitanov},\ and\ \citenamefont {Singer}}]{Ivanov16}%
  \BibitemOpen
  \bibfield  {author} {\bibinfo {author} {\bibfnamefont {P.~A.}\ \bibnamefont
  {Ivanov}}, \bibinfo {author} {\bibfnamefont {N.~V.}\ \bibnamefont {Vitanov}},
  \ and\ \bibinfo {author} {\bibfnamefont {K.}~\bibnamefont {Singer}},\ }\href
  {\doibase 10.1038/srep28078} {\bibfield  {journal} {\bibinfo  {journal}
  {Scientific Reports}\ }\textbf {\bibinfo {volume} {6}},\ \bibinfo {pages}
  {28078} (\bibinfo {year} {2016})}\BibitemShut {NoStop}%
\bibitem [{\citenamefont {Schrinski}\ \emph {et~al.}(2023)\citenamefont
  {Schrinski}, \citenamefont {Yang}, \citenamefont {von L\"upke}, \citenamefont
  {Bild}, \citenamefont {Chu}, \citenamefont {Hornberger}, \citenamefont
  {Nimmrichter},\ and\ \citenamefont {Fadel}}]{macroPRL}%
  \BibitemOpen
  \bibfield  {author} {\bibinfo {author} {\bibfnamefont {B.}~\bibnamefont
  {Schrinski}}, \bibinfo {author} {\bibfnamefont {Y.}~\bibnamefont {Yang}},
  \bibinfo {author} {\bibfnamefont {U.}~\bibnamefont {von L\"upke}}, \bibinfo
  {author} {\bibfnamefont {M.}~\bibnamefont {Bild}}, \bibinfo {author}
  {\bibfnamefont {Y.}~\bibnamefont {Chu}}, \bibinfo {author} {\bibfnamefont
  {K.}~\bibnamefont {Hornberger}}, \bibinfo {author} {\bibfnamefont
  {S.}~\bibnamefont {Nimmrichter}}, \ and\ \bibinfo {author} {\bibfnamefont
  {M.}~\bibnamefont {Fadel}},\ }\href {\doibase 10.1103/PhysRevLett.130.133604}
  {\bibfield  {journal} {\bibinfo  {journal} {Phys. Rev. Lett.}\ }\textbf
  {\bibinfo {volume} {130}},\ \bibinfo {pages} {133604} (\bibinfo {year}
  {2023})}\BibitemShut {NoStop}%
\bibitem [{\citenamefont {Carney}\ \emph {et~al.}(2021)\citenamefont {Carney},
  \citenamefont {Krnjaic}, \citenamefont {Moore}, \citenamefont {Regal},
  \citenamefont {Afek}, \citenamefont {Bhave}, \citenamefont {Brubaker},
  \citenamefont {Corbitt}, \citenamefont {Cripe}, \citenamefont {Crisosto},
  \citenamefont {Geraci}, \citenamefont {Ghosh}, \citenamefont {Harris},
  \citenamefont {Hook}, \citenamefont {Kolb}, \citenamefont {Kunjummen},
  \citenamefont {Lang}, \citenamefont {Li}, \citenamefont {Lin}, \citenamefont
  {Liu}, \citenamefont {Lykken}, \citenamefont {Magrini}, \citenamefont
  {Manley}, \citenamefont {Matsumoto}, \citenamefont {Monte}, \citenamefont
  {Monteiro}, \citenamefont {Purdy}, \citenamefont {Riedel}, \citenamefont
  {Singh}, \citenamefont {Singh}, \citenamefont {Sinha}, \citenamefont
  {Taylor}, \citenamefont {Qin}, \citenamefont {Wilson},\ and\ \citenamefont
  {Zhao}}]{Carney21}%
  \BibitemOpen
  \bibfield  {author} {\bibinfo {author} {\bibfnamefont {D.}~\bibnamefont
  {Carney}}, \bibinfo {author} {\bibfnamefont {G.}~\bibnamefont {Krnjaic}},
  \bibinfo {author} {\bibfnamefont {D.~C.}\ \bibnamefont {Moore}}, \bibinfo
  {author} {\bibfnamefont {C.~A.}\ \bibnamefont {Regal}}, \bibinfo {author}
  {\bibfnamefont {G.}~\bibnamefont {Afek}}, \bibinfo {author} {\bibfnamefont
  {S.}~\bibnamefont {Bhave}}, \bibinfo {author} {\bibfnamefont
  {B.}~\bibnamefont {Brubaker}}, \bibinfo {author} {\bibfnamefont
  {T.}~\bibnamefont {Corbitt}}, \bibinfo {author} {\bibfnamefont
  {J.}~\bibnamefont {Cripe}}, \bibinfo {author} {\bibfnamefont
  {N.}~\bibnamefont {Crisosto}}, \bibinfo {author} {\bibfnamefont
  {A.}~\bibnamefont {Geraci}}, \bibinfo {author} {\bibfnamefont
  {S.}~\bibnamefont {Ghosh}}, \bibinfo {author} {\bibfnamefont {J.~G.~E.}\
  \bibnamefont {Harris}}, \bibinfo {author} {\bibfnamefont {A.}~\bibnamefont
  {Hook}}, \bibinfo {author} {\bibfnamefont {E.~W.}\ \bibnamefont {Kolb}},
  \bibinfo {author} {\bibfnamefont {J.}~\bibnamefont {Kunjummen}}, \bibinfo
  {author} {\bibfnamefont {R.~F.}\ \bibnamefont {Lang}}, \bibinfo {author}
  {\bibfnamefont {T.}~\bibnamefont {Li}}, \bibinfo {author} {\bibfnamefont
  {T.}~\bibnamefont {Lin}}, \bibinfo {author} {\bibfnamefont {Z.}~\bibnamefont
  {Liu}}, \bibinfo {author} {\bibfnamefont {J.}~\bibnamefont {Lykken}},
  \bibinfo {author} {\bibfnamefont {L.}~\bibnamefont {Magrini}}, \bibinfo
  {author} {\bibfnamefont {J.}~\bibnamefont {Manley}}, \bibinfo {author}
  {\bibfnamefont {N.}~\bibnamefont {Matsumoto}}, \bibinfo {author}
  {\bibfnamefont {A.}~\bibnamefont {Monte}}, \bibinfo {author} {\bibfnamefont
  {F.}~\bibnamefont {Monteiro}}, \bibinfo {author} {\bibfnamefont
  {T.}~\bibnamefont {Purdy}}, \bibinfo {author} {\bibfnamefont {C.~J.}\
  \bibnamefont {Riedel}}, \bibinfo {author} {\bibfnamefont {R.}~\bibnamefont
  {Singh}}, \bibinfo {author} {\bibfnamefont {S.}~\bibnamefont {Singh}},
  \bibinfo {author} {\bibfnamefont {K.}~\bibnamefont {Sinha}}, \bibinfo
  {author} {\bibfnamefont {J.~M.}\ \bibnamefont {Taylor}}, \bibinfo {author}
  {\bibfnamefont {J.}~\bibnamefont {Qin}}, \bibinfo {author} {\bibfnamefont
  {D.~J.}\ \bibnamefont {Wilson}}, \ and\ \bibinfo {author} {\bibfnamefont
  {Y.}~\bibnamefont {Zhao}},\ }\href {\doibase 10.1088/2058-9565/abcfcd}
  {\bibfield  {journal} {\bibinfo  {journal} {Quantum Science and Technology}\
  }\textbf {\bibinfo {volume} {6}},\ \bibinfo {pages} {024002} (\bibinfo {year}
  {2021})}\BibitemShut {NoStop}%
\bibitem [{\citenamefont {Bonaldi}\ \emph {et~al.}(2020)\citenamefont
  {Bonaldi}, \citenamefont {Borrielli}, \citenamefont {Chowdhury},
  \citenamefont {Di~Giuseppe}, \citenamefont {Li}, \citenamefont {Malossi},
  \citenamefont {Marino}, \citenamefont {Morana}, \citenamefont {Natali},
  \citenamefont {Piergentili}, \citenamefont {Prodi}, \citenamefont {Sarro},
  \citenamefont {Serra}, \citenamefont {Vezio}, \citenamefont {Vitali},\ and\
  \citenamefont {Marin}}]{BonaldiQGtest}%
  \BibitemOpen
  \bibfield  {author} {\bibinfo {author} {\bibfnamefont {M.}~\bibnamefont
  {Bonaldi}}, \bibinfo {author} {\bibfnamefont {A.}~\bibnamefont {Borrielli}},
  \bibinfo {author} {\bibfnamefont {A.}~\bibnamefont {Chowdhury}}, \bibinfo
  {author} {\bibfnamefont {G.}~\bibnamefont {Di~Giuseppe}}, \bibinfo {author}
  {\bibfnamefont {W.}~\bibnamefont {Li}}, \bibinfo {author} {\bibfnamefont
  {N.}~\bibnamefont {Malossi}}, \bibinfo {author} {\bibfnamefont
  {F.}~\bibnamefont {Marino}}, \bibinfo {author} {\bibfnamefont
  {B.}~\bibnamefont {Morana}}, \bibinfo {author} {\bibfnamefont
  {R.}~\bibnamefont {Natali}}, \bibinfo {author} {\bibfnamefont
  {P.}~\bibnamefont {Piergentili}}, \bibinfo {author} {\bibfnamefont {G.~A.}\
  \bibnamefont {Prodi}}, \bibinfo {author} {\bibfnamefont {P.~M.}\ \bibnamefont
  {Sarro}}, \bibinfo {author} {\bibfnamefont {E.}~\bibnamefont {Serra}},
  \bibinfo {author} {\bibfnamefont {P.}~\bibnamefont {Vezio}}, \bibinfo
  {author} {\bibfnamefont {D.}~\bibnamefont {Vitali}}, \ and\ \bibinfo {author}
  {\bibfnamefont {F.}~\bibnamefont {Marin}},\ }\href {\doibase
  10.1140/epjd/e2020-10184-6} {\bibfield  {journal} {\bibinfo  {journal} {The
  European Physical Journal D}\ }\textbf {\bibinfo {volume} {74}},\ \bibinfo
  {pages} {178} (\bibinfo {year} {2020})}\BibitemShut {NoStop}%
\bibitem [{\citenamefont {Clerk}\ \emph {et~al.}(2020)\citenamefont {Clerk},
  \citenamefont {Lehnert}, \citenamefont {Bertet}, \citenamefont {Petta},\ and\
  \citenamefont {Nakamura}}]{ClerkNatReview}%
  \BibitemOpen
  \bibfield  {author} {\bibinfo {author} {\bibfnamefont {A.~A.}\ \bibnamefont
  {Clerk}}, \bibinfo {author} {\bibfnamefont {K.~W.}\ \bibnamefont {Lehnert}},
  \bibinfo {author} {\bibfnamefont {P.}~\bibnamefont {Bertet}}, \bibinfo
  {author} {\bibfnamefont {J.~R.}\ \bibnamefont {Petta}}, \ and\ \bibinfo
  {author} {\bibfnamefont {Y.}~\bibnamefont {Nakamura}},\ }\href {\doibase
  10.1038/s41567-020-0797-9} {\bibfield  {journal} {\bibinfo  {journal} {Nature
  Physics}\ }\textbf {\bibinfo {volume} {16}},\ \bibinfo {pages} {257}
  (\bibinfo {year} {2020})}\BibitemShut {NoStop}%
\bibitem [{\citenamefont {Chu}\ and\ \citenamefont
  {Gr\"oblacher}(2020)}]{Chu20}%
  \BibitemOpen
  \bibfield  {author} {\bibinfo {author} {\bibfnamefont {Y.}~\bibnamefont
  {Chu}}\ and\ \bibinfo {author} {\bibfnamefont {S.}~\bibnamefont
  {Gr\"oblacher}},\ }\href {\doibase http://dx.doi.org/10.1063/5.0021088}
  {\bibfield  {journal} {\bibinfo  {journal} {Applied Physics Letters}\
  }\textbf {\bibinfo {volume} {117}},\ \bibinfo {pages} {150503} (\bibinfo
  {year} {2020})}\BibitemShut {NoStop}%
\bibitem [{\citenamefont {Satzinger}\ \emph {et~al.}(2018)\citenamefont
  {Satzinger}, \citenamefont {Zhong}, \citenamefont {Chang}, \citenamefont
  {Peairs}, \citenamefont {Bienfait}, \citenamefont {Chou}, \citenamefont
  {Cleland}, \citenamefont {Conner}, \citenamefont {Dumur}, \citenamefont
  {Grebel}, \citenamefont {Gutierrez}, \citenamefont {November}, \citenamefont
  {Povey}, \citenamefont {Whiteley}, \citenamefont {Awschalom}, \citenamefont
  {Schuster},\ and\ \citenamefont {Cleland}}]{Satzinger18}%
  \BibitemOpen
  \bibfield  {author} {\bibinfo {author} {\bibfnamefont {K.~J.}\ \bibnamefont
  {Satzinger}}, \bibinfo {author} {\bibfnamefont {Y.~P.}\ \bibnamefont
  {Zhong}}, \bibinfo {author} {\bibfnamefont {H.-S.}\ \bibnamefont {Chang}},
  \bibinfo {author} {\bibfnamefont {G.~A.}\ \bibnamefont {Peairs}}, \bibinfo
  {author} {\bibfnamefont {A.}~\bibnamefont {Bienfait}}, \bibinfo {author}
  {\bibfnamefont {M.-H.}\ \bibnamefont {Chou}}, \bibinfo {author}
  {\bibfnamefont {A.~Y.}\ \bibnamefont {Cleland}}, \bibinfo {author}
  {\bibfnamefont {C.~R.}\ \bibnamefont {Conner}}, \bibinfo {author}
  {\bibfnamefont {E.}~\bibnamefont {Dumur}}, \bibinfo {author} {\bibfnamefont
  {J.}~\bibnamefont {Grebel}}, \bibinfo {author} {\bibfnamefont
  {I.}~\bibnamefont {Gutierrez}}, \bibinfo {author} {\bibfnamefont {B.~H.}\
  \bibnamefont {November}}, \bibinfo {author} {\bibfnamefont {R.~G.}\
  \bibnamefont {Povey}}, \bibinfo {author} {\bibfnamefont {S.~J.}\ \bibnamefont
  {Whiteley}}, \bibinfo {author} {\bibfnamefont {D.~D.}\ \bibnamefont
  {Awschalom}}, \bibinfo {author} {\bibfnamefont {D.~I.}\ \bibnamefont
  {Schuster}}, \ and\ \bibinfo {author} {\bibfnamefont {A.~N.}\ \bibnamefont
  {Cleland}},\ }\href {\doibase http://dx.doi.org/10.1038/s41586-018-0719-5}
  {\bibfield  {journal} {\bibinfo  {journal} {Nature}\ }\textbf {\bibinfo
  {volume} {563}},\ \bibinfo {pages} {661} (\bibinfo {year}
  {2018})}\BibitemShut {NoStop}%
\bibitem [{\citenamefont {Arrangoiz-Arriola}\ \emph {et~al.}(2018)\citenamefont
  {Arrangoiz-Arriola}, \citenamefont {Wollack}, \citenamefont {Pechal},
  \citenamefont {Witmer}, \citenamefont {Hill},\ and\ \citenamefont
  {Safavi-Naeini}}]{Arrangoiz-Arriola2018}%
  \BibitemOpen
  \bibfield  {author} {\bibinfo {author} {\bibfnamefont {P.}~\bibnamefont
  {Arrangoiz-Arriola}}, \bibinfo {author} {\bibfnamefont {E.~A.}\ \bibnamefont
  {Wollack}}, \bibinfo {author} {\bibfnamefont {M.}~\bibnamefont {Pechal}},
  \bibinfo {author} {\bibfnamefont {J.~D.}\ \bibnamefont {Witmer}}, \bibinfo
  {author} {\bibfnamefont {J.~T.}\ \bibnamefont {Hill}}, \ and\ \bibinfo
  {author} {\bibfnamefont {A.~H.}\ \bibnamefont {Safavi-Naeini}},\ }\href
  {\doibase 10.1103/PhysRevX.8.031007} {\bibfield  {journal} {\bibinfo
  {journal} {Phys. Rev. X}\ }\textbf {\bibinfo {volume} {8}},\ \bibinfo {pages}
  {031007} (\bibinfo {year} {2018})}\BibitemShut {NoStop}%
\bibitem [{\citenamefont {Chu}\ \emph {et~al.}(2018)\citenamefont {Chu},
  \citenamefont {Kharel}, \citenamefont {Yoon}, \citenamefont {Frunzio},
  \citenamefont {Rakich},\ and\ \citenamefont {Schoelkopf}}]{Chu2018}%
  \BibitemOpen
  \bibfield  {author} {\bibinfo {author} {\bibfnamefont {Y.}~\bibnamefont
  {Chu}}, \bibinfo {author} {\bibfnamefont {P.}~\bibnamefont {Kharel}},
  \bibinfo {author} {\bibfnamefont {T.}~\bibnamefont {Yoon}}, \bibinfo {author}
  {\bibfnamefont {L.}~\bibnamefont {Frunzio}}, \bibinfo {author} {\bibfnamefont
  {P.~T.}\ \bibnamefont {Rakich}}, \ and\ \bibinfo {author} {\bibfnamefont
  {R.~J.}\ \bibnamefont {Schoelkopf}},\ }\href
  {https://doi.org/10.1038/s41586-018-0717-7} {\bibfield  {journal} {\bibinfo
  {journal} {Nature}\ }\textbf {\bibinfo {volume} {563}},\ \bibinfo {pages}
  {666} (\bibinfo {year} {2018})}\BibitemShut {NoStop}%
\bibitem [{\citenamefont {Bild}\ \emph {et~al.}(2023)\citenamefont {Bild},
  \citenamefont {Fadel}, \citenamefont {Yang}, \citenamefont {von L\"upke},
  \citenamefont {Martin}, \citenamefont {Bruno},\ and\ \citenamefont
  {Chu}}]{catSCI23}%
  \BibitemOpen
  \bibfield  {author} {\bibinfo {author} {\bibfnamefont {M.}~\bibnamefont
  {Bild}}, \bibinfo {author} {\bibfnamefont {M.}~\bibnamefont {Fadel}},
  \bibinfo {author} {\bibfnamefont {Y.}~\bibnamefont {Yang}}, \bibinfo {author}
  {\bibfnamefont {U.}~\bibnamefont {von L\"upke}}, \bibinfo {author}
  {\bibfnamefont {P.}~\bibnamefont {Martin}}, \bibinfo {author} {\bibfnamefont
  {A.}~\bibnamefont {Bruno}}, \ and\ \bibinfo {author} {\bibfnamefont
  {Y.}~\bibnamefont {Chu}},\ }\href {\doibase 10.1126/science.adf7553}
  {\bibfield  {journal} {\bibinfo  {journal} {Science}\ }\textbf {\bibinfo
  {volume} {380}},\ \bibinfo {pages} {274} (\bibinfo {year}
  {2023})}\BibitemShut {NoStop}%
\bibitem [{\citenamefont {Lloyd}\ and\ \citenamefont
  {Braunstein}(1999)}]{LloydCVPRL19}%
  \BibitemOpen
  \bibfield  {author} {\bibinfo {author} {\bibfnamefont {S.}~\bibnamefont
  {Lloyd}}\ and\ \bibinfo {author} {\bibfnamefont {S.~L.}\ \bibnamefont
  {Braunstein}},\ }\href {\doibase 10.1103/PhysRevLett.82.1784} {\bibfield
  {journal} {\bibinfo  {journal} {Phys. Rev. Lett.}\ }\textbf {\bibinfo
  {volume} {82}},\ \bibinfo {pages} {1784} (\bibinfo {year}
  {1999})}\BibitemShut {NoStop}%
\bibitem [{\citenamefont {Mari}\ and\ \citenamefont
  {Eisert}(2012)}]{MariPRL12}%
  \BibitemOpen
  \bibfield  {author} {\bibinfo {author} {\bibfnamefont {A.}~\bibnamefont
  {Mari}}\ and\ \bibinfo {author} {\bibfnamefont {J.}~\bibnamefont {Eisert}},\
  }\href {\doibase 10.1103/PhysRevLett.109.230503} {\bibfield  {journal}
  {\bibinfo  {journal} {Phys. Rev. Lett.}\ }\textbf {\bibinfo {volume} {109}},\
  \bibinfo {pages} {230503} (\bibinfo {year} {2012})}\BibitemShut {NoStop}%
\bibitem [{\citenamefont {Braunstein}\ and\ \citenamefont {van
  Loock}(2005)}]{BraunsteinRMP}%
  \BibitemOpen
  \bibfield  {author} {\bibinfo {author} {\bibfnamefont {S.~L.}\ \bibnamefont
  {Braunstein}}\ and\ \bibinfo {author} {\bibfnamefont {P.}~\bibnamefont {van
  Loock}},\ }\href {\doibase 10.1103/RevModPhys.77.513} {\bibfield  {journal}
  {\bibinfo  {journal} {Rev. Mod. Phys.}\ }\textbf {\bibinfo {volume} {77}},\
  \bibinfo {pages} {513} (\bibinfo {year} {2005})}\BibitemShut {NoStop}%
\bibitem [{\citenamefont {Weedbrook}\ \emph {et~al.}(2012)\citenamefont
  {Weedbrook}, \citenamefont {Pirandola}, \citenamefont {Garc\'{\i}a-Patr\'on},
  \citenamefont {Cerf}, \citenamefont {Ralph}, \citenamefont {Shapiro},\ and\
  \citenamefont {Lloyd}}]{GaussQinfoRMP12}%
  \BibitemOpen
  \bibfield  {author} {\bibinfo {author} {\bibfnamefont {C.}~\bibnamefont
  {Weedbrook}}, \bibinfo {author} {\bibfnamefont {S.}~\bibnamefont
  {Pirandola}}, \bibinfo {author} {\bibfnamefont {R.}~\bibnamefont
  {Garc\'{\i}a-Patr\'on}}, \bibinfo {author} {\bibfnamefont {N.~J.}\
  \bibnamefont {Cerf}}, \bibinfo {author} {\bibfnamefont {T.~C.}\ \bibnamefont
  {Ralph}}, \bibinfo {author} {\bibfnamefont {J.~H.}\ \bibnamefont {Shapiro}},
  \ and\ \bibinfo {author} {\bibfnamefont {S.}~\bibnamefont {Lloyd}},\ }\href
  {\doibase 10.1103/RevModPhys.84.621} {\bibfield  {journal} {\bibinfo
  {journal} {Rev. Mod. Phys.}\ }\textbf {\bibinfo {volume} {84}},\ \bibinfo
  {pages} {621} (\bibinfo {year} {2012})}\BibitemShut {NoStop}%
\bibitem [{\citenamefont {von L\"{u}pke}\ \emph {et~al.}(2023)\citenamefont
  {von L\"{u}pke}, \citenamefont {Rodrigues}, \citenamefont {Yang},
  \citenamefont {Fadel},\ and\ \citenamefont {Chu}}]{BSpaper}%
  \BibitemOpen
  \bibfield  {author} {\bibinfo {author} {\bibfnamefont {U.}~\bibnamefont {von
  L\"{u}pke}}, \bibinfo {author} {\bibfnamefont {I.~C.}\ \bibnamefont
  {Rodrigues}}, \bibinfo {author} {\bibfnamefont {Y.}~\bibnamefont {Yang}},
  \bibinfo {author} {\bibfnamefont {M.}~\bibnamefont {Fadel}}, \ and\ \bibinfo
  {author} {\bibfnamefont {Y.}~\bibnamefont {Chu}},\ }\href@noop {} {\enquote
  {\bibinfo {title} {Engineering phonon-phonon interactions in multimode
  circuit quantum acousto-dynamics},}\ } (\bibinfo {year} {2023}),\ \Eprint
  {http://arxiv.org/abs/2303.00730} {arXiv:2303.00730 [quant-ph]} \BibitemShut
  {NoStop}%
\bibitem [{\citenamefont {Meekhof}\ \emph {et~al.}(1996)\citenamefont
  {Meekhof}, \citenamefont {Monroe}, \citenamefont {King}, \citenamefont
  {Itano},\ and\ \citenamefont {Wineland}}]{MeekhofPRL96}%
  \BibitemOpen
  \bibfield  {author} {\bibinfo {author} {\bibfnamefont {D.~M.}\ \bibnamefont
  {Meekhof}}, \bibinfo {author} {\bibfnamefont {C.}~\bibnamefont {Monroe}},
  \bibinfo {author} {\bibfnamefont {B.~E.}\ \bibnamefont {King}}, \bibinfo
  {author} {\bibfnamefont {W.~M.}\ \bibnamefont {Itano}}, \ and\ \bibinfo
  {author} {\bibfnamefont {D.~J.}\ \bibnamefont {Wineland}},\ }\href {\doibase
  10.1103/PhysRevLett.76.1796} {\bibfield  {journal} {\bibinfo  {journal}
  {Phys. Rev. Lett.}\ }\textbf {\bibinfo {volume} {76}},\ \bibinfo {pages}
  {1796} (\bibinfo {year} {1996})}\BibitemShut {NoStop}%
\bibitem [{\citenamefont {Wollman}\ \emph {et~al.}(2015)\citenamefont
  {Wollman}, \citenamefont {Lei}, \citenamefont {Weinstein}, \citenamefont
  {Suh}, \citenamefont {Kronwald}, \citenamefont {Marquardt}, \citenamefont
  {Clerk},\ and\ \citenamefont {Schwab}}]{WollmanSCI15}%
  \BibitemOpen
  \bibfield  {author} {\bibinfo {author} {\bibfnamefont {E.~E.}\ \bibnamefont
  {Wollman}}, \bibinfo {author} {\bibfnamefont {C.~U.}\ \bibnamefont {Lei}},
  \bibinfo {author} {\bibfnamefont {A.~J.}\ \bibnamefont {Weinstein}}, \bibinfo
  {author} {\bibfnamefont {J.}~\bibnamefont {Suh}}, \bibinfo {author}
  {\bibfnamefont {A.}~\bibnamefont {Kronwald}}, \bibinfo {author}
  {\bibfnamefont {F.}~\bibnamefont {Marquardt}}, \bibinfo {author}
  {\bibfnamefont {A.~A.}\ \bibnamefont {Clerk}}, \ and\ \bibinfo {author}
  {\bibfnamefont {K.~C.}\ \bibnamefont {Schwab}},\ }\href {\doibase
  10.1126/science.aac5138} {\bibfield  {journal} {\bibinfo  {journal}
  {Science}\ }\textbf {\bibinfo {volume} {349}},\ \bibinfo {pages} {952}
  (\bibinfo {year} {2015})}\BibitemShut {NoStop}%
\bibitem [{\citenamefont {Pirkkalainen}\ \emph {et~al.}(2015)\citenamefont
  {Pirkkalainen}, \citenamefont {Damsk\"agg}, \citenamefont {Brandt},
  \citenamefont {Massel},\ and\ \citenamefont
  {Sillanp\"a\"a}}]{PirkkalainenPRL15}%
  \BibitemOpen
  \bibfield  {author} {\bibinfo {author} {\bibfnamefont {J.-M.}\ \bibnamefont
  {Pirkkalainen}}, \bibinfo {author} {\bibfnamefont {E.}~\bibnamefont
  {Damsk\"agg}}, \bibinfo {author} {\bibfnamefont {M.}~\bibnamefont {Brandt}},
  \bibinfo {author} {\bibfnamefont {F.}~\bibnamefont {Massel}}, \ and\ \bibinfo
  {author} {\bibfnamefont {M.~A.}\ \bibnamefont {Sillanp\"a\"a}},\ }\href
  {\doibase 10.1103/PhysRevLett.115.243601} {\bibfield  {journal} {\bibinfo
  {journal} {Phys. Rev. Lett.}\ }\textbf {\bibinfo {volume} {115}},\ \bibinfo
  {pages} {243601} (\bibinfo {year} {2015})}\BibitemShut {NoStop}%
\bibitem [{\citenamefont {Lecocq}\ \emph {et~al.}(2015)\citenamefont {Lecocq},
  \citenamefont {Clark}, \citenamefont {Simmonds}, \citenamefont {Aumentado},\
  and\ \citenamefont {Teufel}}]{LecocqPRX15}%
  \BibitemOpen
  \bibfield  {author} {\bibinfo {author} {\bibfnamefont {F.}~\bibnamefont
  {Lecocq}}, \bibinfo {author} {\bibfnamefont {J.~B.}\ \bibnamefont {Clark}},
  \bibinfo {author} {\bibfnamefont {R.~W.}\ \bibnamefont {Simmonds}}, \bibinfo
  {author} {\bibfnamefont {J.}~\bibnamefont {Aumentado}}, \ and\ \bibinfo
  {author} {\bibfnamefont {J.~D.}\ \bibnamefont {Teufel}},\ }\href {\doibase
  10.1103/PhysRevX.5.041037} {\bibfield  {journal} {\bibinfo  {journal} {Phys.
  Rev. X}\ }\textbf {\bibinfo {volume} {5}},\ \bibinfo {pages} {041037}
  (\bibinfo {year} {2015})}\BibitemShut {NoStop}%
\bibitem [{\citenamefont {Delaney}\ \emph {et~al.}(2019)\citenamefont
  {Delaney}, \citenamefont {Reed}, \citenamefont {Andrews},\ and\ \citenamefont
  {Lehnert}}]{DelaneyPRL19}%
  \BibitemOpen
  \bibfield  {author} {\bibinfo {author} {\bibfnamefont {R.~D.}\ \bibnamefont
  {Delaney}}, \bibinfo {author} {\bibfnamefont {A.~P.}\ \bibnamefont {Reed}},
  \bibinfo {author} {\bibfnamefont {R.~W.}\ \bibnamefont {Andrews}}, \ and\
  \bibinfo {author} {\bibfnamefont {K.~W.}\ \bibnamefont {Lehnert}},\ }\href
  {\doibase 10.1103/PhysRevLett.123.183603} {\bibfield  {journal} {\bibinfo
  {journal} {Phys. Rev. Lett.}\ }\textbf {\bibinfo {volume} {123}},\ \bibinfo
  {pages} {183603} (\bibinfo {year} {2019})}\BibitemShut {NoStop}%
\bibitem [{\citenamefont {Youssefi}\ \emph {et~al.}(2023)\citenamefont
  {Youssefi}, \citenamefont {Kono}, \citenamefont {Chegnizadeh},\ and\
  \citenamefont {Kippenberg}}]{youssefi2023squeezed}%
  \BibitemOpen
  \bibfield  {author} {\bibinfo {author} {\bibfnamefont {A.}~\bibnamefont
  {Youssefi}}, \bibinfo {author} {\bibfnamefont {S.}~\bibnamefont {Kono}},
  \bibinfo {author} {\bibfnamefont {M.}~\bibnamefont {Chegnizadeh}}, \ and\
  \bibinfo {author} {\bibfnamefont {T.~J.}\ \bibnamefont {Kippenberg}},\ }\href
  {\doibase 10.1038/s41567-023-02135-y} {\bibfield  {journal} {\bibinfo
  {journal} {Nature Physics}\ }\textbf {\bibinfo {volume} {19}},\ \bibinfo
  {pages} {1697} (\bibinfo {year} {2023})}\BibitemShut {NoStop}%
\bibitem [{\citenamefont {Andersson}\ \emph {et~al.}(2022)\citenamefont
  {Andersson}, \citenamefont {Jolin}, \citenamefont {Scigliuzzo}, \citenamefont
  {Borgani}, \citenamefont {Thol\'en}, \citenamefont {Rivera~Hern\'andez},
  \citenamefont {Shumeiko}, \citenamefont {Haviland},\ and\ \citenamefont
  {Delsing}}]{SAWsqPRXQ22}%
  \BibitemOpen
  \bibfield  {author} {\bibinfo {author} {\bibfnamefont {G.}~\bibnamefont
  {Andersson}}, \bibinfo {author} {\bibfnamefont {S.~W.}\ \bibnamefont
  {Jolin}}, \bibinfo {author} {\bibfnamefont {M.}~\bibnamefont {Scigliuzzo}},
  \bibinfo {author} {\bibfnamefont {R.}~\bibnamefont {Borgani}}, \bibinfo
  {author} {\bibfnamefont {M.~O.}\ \bibnamefont {Thol\'en}}, \bibinfo {author}
  {\bibfnamefont {J.}~\bibnamefont {Rivera~Hern\'andez}}, \bibinfo {author}
  {\bibfnamefont {V.}~\bibnamefont {Shumeiko}}, \bibinfo {author}
  {\bibfnamefont {D.~B.}\ \bibnamefont {Haviland}}, \ and\ \bibinfo {author}
  {\bibfnamefont {P.}~\bibnamefont {Delsing}},\ }\href {\doibase
  10.1103/PRXQuantum.3.010312} {\bibfield  {journal} {\bibinfo  {journal} {PRX
  Quantum}\ }\textbf {\bibinfo {volume} {3}},\ \bibinfo {pages} {010312}
  (\bibinfo {year} {2022})}\BibitemShut {NoStop}%
\bibitem [{\citenamefont {Samanta}\ \emph {et~al.}(2023)\citenamefont
  {Samanta}, \citenamefont {De~Bonis}, \citenamefont {M{\o}ller}, \citenamefont
  {Tormo-Queralt}, \citenamefont {Yang}, \citenamefont {Urgell}, \citenamefont
  {Stamenic}, \citenamefont {Thibeault}, \citenamefont {Jin}, \citenamefont
  {Czaplewski}, \citenamefont {Pistolesi},\ and\ \citenamefont
  {Bachtold}}]{SamantaNatPhys}%
  \BibitemOpen
  \bibfield  {author} {\bibinfo {author} {\bibfnamefont {C.}~\bibnamefont
  {Samanta}}, \bibinfo {author} {\bibfnamefont {S.~L.}\ \bibnamefont
  {De~Bonis}}, \bibinfo {author} {\bibfnamefont {C.~B.}\ \bibnamefont
  {M{\o}ller}}, \bibinfo {author} {\bibfnamefont {R.}~\bibnamefont
  {Tormo-Queralt}}, \bibinfo {author} {\bibfnamefont {W.}~\bibnamefont {Yang}},
  \bibinfo {author} {\bibfnamefont {C.}~\bibnamefont {Urgell}}, \bibinfo
  {author} {\bibfnamefont {B.}~\bibnamefont {Stamenic}}, \bibinfo {author}
  {\bibfnamefont {B.}~\bibnamefont {Thibeault}}, \bibinfo {author}
  {\bibfnamefont {Y.}~\bibnamefont {Jin}}, \bibinfo {author} {\bibfnamefont
  {D.~A.}\ \bibnamefont {Czaplewski}}, \bibinfo {author} {\bibfnamefont
  {F.}~\bibnamefont {Pistolesi}}, \ and\ \bibinfo {author} {\bibfnamefont
  {A.}~\bibnamefont {Bachtold}},\ }\href {\doibase 10.1038/s41567-023-02065-9}
  {\bibfield  {journal} {\bibinfo  {journal} {Nature Physics}\ }\textbf
  {\bibinfo {volume} {19}},\ \bibinfo {pages} {1340} (\bibinfo {year}
  {2023})}\BibitemShut {NoStop}%
\bibitem [{\citenamefont {von L{\"u}pke}\ \emph {et~al.}(2022)\citenamefont
  {von L{\"u}pke}, \citenamefont {Yang}, \citenamefont {Bild}, \citenamefont
  {Michaud}, \citenamefont {Fadel},\ and\ \citenamefont {Chu}}]{vonLupke22}%
  \BibitemOpen
  \bibfield  {author} {\bibinfo {author} {\bibfnamefont {U.}~\bibnamefont {von
  L{\"u}pke}}, \bibinfo {author} {\bibfnamefont {Y.}~\bibnamefont {Yang}},
  \bibinfo {author} {\bibfnamefont {M.}~\bibnamefont {Bild}}, \bibinfo {author}
  {\bibfnamefont {L.}~\bibnamefont {Michaud}}, \bibinfo {author} {\bibfnamefont
  {M.}~\bibnamefont {Fadel}}, \ and\ \bibinfo {author} {\bibfnamefont
  {Y.}~\bibnamefont {Chu}},\ }\href {\doibase 10.1038/s41567-022-01591-2}
  {\bibfield  {journal} {\bibinfo  {journal} {Nature Physics}\ }\textbf
  {\bibinfo {volume} {18}},\ \bibinfo {pages} {794} (\bibinfo {year}
  {2022})}\BibitemShut {NoStop}%
\bibitem [{SM()}]{SM}%
  \BibitemOpen
  \href@noop {} {\bibinfo  {journal} {See supplementary materials}\
  }\BibitemShut {NoStop}%
\bibitem [{\citenamefont {Zhang}\ \emph {et~al.}(2019)\citenamefont {Zhang},
  \citenamefont {Lester}, \citenamefont {Gao}, \citenamefont {Jiang},
  \citenamefont {Schoelkopf},\ and\ \citenamefont
  {Girvin}}]{zhang2019engineering}%
  \BibitemOpen
\bibfield  {journal} {  }\bibfield  {author} {\bibinfo {author} {\bibfnamefont
  {Y.}~\bibnamefont {Zhang}}, \bibinfo {author} {\bibfnamefont {B.~J.}\
  \bibnamefont {Lester}}, \bibinfo {author} {\bibfnamefont {Y.~Y.}\
  \bibnamefont {Gao}}, \bibinfo {author} {\bibfnamefont {L.}~\bibnamefont
  {Jiang}}, \bibinfo {author} {\bibfnamefont {R.}~\bibnamefont {Schoelkopf}}, \
  and\ \bibinfo {author} {\bibfnamefont {S.}~\bibnamefont {Girvin}},\ }\href
  {\doibase http://dx.doi.org/10.1103/PhysRevA.99.012314} {\bibfield  {journal}
  {\bibinfo  {journal} {Physical Review A}\ }\textbf {\bibinfo {volume} {99}},\
  \bibinfo {pages} {012314} (\bibinfo {year} {2019})}\BibitemShut {NoStop}%
\bibitem [{\citenamefont {Wang}\ \emph {et~al.}(2020)\citenamefont {Wang},
  \citenamefont {Curtis}, \citenamefont {Lester}, \citenamefont {Zhang},
  \citenamefont {Gao}, \citenamefont {Freeze}, \citenamefont {Batista},
  \citenamefont {Vaccaro}, \citenamefont {Chuang}, \citenamefont {Frunzio},
  \citenamefont {Jiang}, \citenamefont {Girvin},\ and\ \citenamefont
  {Schoelkopf}}]{Wang2019}%
  \BibitemOpen
  \bibfield  {author} {\bibinfo {author} {\bibfnamefont {C.~S.}\ \bibnamefont
  {Wang}}, \bibinfo {author} {\bibfnamefont {J.~C.}\ \bibnamefont {Curtis}},
  \bibinfo {author} {\bibfnamefont {B.~J.}\ \bibnamefont {Lester}}, \bibinfo
  {author} {\bibfnamefont {Y.}~\bibnamefont {Zhang}}, \bibinfo {author}
  {\bibfnamefont {Y.~Y.}\ \bibnamefont {Gao}}, \bibinfo {author} {\bibfnamefont
  {J.}~\bibnamefont {Freeze}}, \bibinfo {author} {\bibfnamefont {V.~S.}\
  \bibnamefont {Batista}}, \bibinfo {author} {\bibfnamefont {P.~H.}\
  \bibnamefont {Vaccaro}}, \bibinfo {author} {\bibfnamefont {I.~L.}\
  \bibnamefont {Chuang}}, \bibinfo {author} {\bibfnamefont {L.}~\bibnamefont
  {Frunzio}}, \bibinfo {author} {\bibfnamefont {L.}~\bibnamefont {Jiang}},
  \bibinfo {author} {\bibfnamefont {S.~M.}\ \bibnamefont {Girvin}}, \ and\
  \bibinfo {author} {\bibfnamefont {R.~J.}\ \bibnamefont {Schoelkopf}},\ }\href
  {\doibase 10.1103/PhysRevX.10.021060} {\bibfield  {journal} {\bibinfo
  {journal} {Phys. Rev. X}\ }\textbf {\bibinfo {volume} {10}},\ \bibinfo
  {pages} {021060} (\bibinfo {year} {2020})}\BibitemShut {NoStop}%
\bibitem [{\citenamefont {Milburn}\ and\ \citenamefont
  {Holmes}(1991)}]{MilburnPRA91}%
  \BibitemOpen
  \bibfield  {author} {\bibinfo {author} {\bibfnamefont {G.~J.}\ \bibnamefont
  {Milburn}}\ and\ \bibinfo {author} {\bibfnamefont {C.~A.}\ \bibnamefont
  {Holmes}},\ }\href {\doibase 10.1103/PhysRevA.44.4704} {\bibfield  {journal}
  {\bibinfo  {journal} {Phys. Rev. A}\ }\textbf {\bibinfo {volume} {44}},\
  \bibinfo {pages} {4704} (\bibinfo {year} {1991})}\BibitemShut {NoStop}%
\bibitem [{\citenamefont {Ch{\'a}vez-Carlos}\ \emph {et~al.}(2023)\citenamefont
  {Ch{\'a}vez-Carlos}, \citenamefont {Lezama}, \citenamefont {Corti{\~n}as},
  \citenamefont {Venkatraman}, \citenamefont {Devoret}, \citenamefont
  {Batista}, \citenamefont {P{\'e}rez-Bernal},\ and\ \citenamefont
  {Santos}}]{CCarlosNPJ23}%
  \BibitemOpen
  \bibfield  {author} {\bibinfo {author} {\bibfnamefont {J.}~\bibnamefont
  {Ch{\'a}vez-Carlos}}, \bibinfo {author} {\bibfnamefont {T.~L.~M.}\
  \bibnamefont {Lezama}}, \bibinfo {author} {\bibfnamefont {R.~G.}\
  \bibnamefont {Corti{\~n}as}}, \bibinfo {author} {\bibfnamefont
  {J.}~\bibnamefont {Venkatraman}}, \bibinfo {author} {\bibfnamefont {M.~H.}\
  \bibnamefont {Devoret}}, \bibinfo {author} {\bibfnamefont {V.~S.}\
  \bibnamefont {Batista}}, \bibinfo {author} {\bibfnamefont {F.}~\bibnamefont
  {P{\'e}rez-Bernal}}, \ and\ \bibinfo {author} {\bibfnamefont {L.~F.}\
  \bibnamefont {Santos}},\ }\href {\doibase 10.1038/s41534-023-00745-1}
  {\bibfield  {journal} {\bibinfo  {journal} {npj Quantum Information}\
  }\textbf {\bibinfo {volume} {9}},\ \bibinfo {pages} {76} (\bibinfo {year}
  {2023})}\BibitemShut {NoStop}%
\bibitem [{\citenamefont {Wielinga}\ and\ \citenamefont
  {Milburn}(1993)}]{WielingaPRA93}%
  \BibitemOpen
  \bibfield  {author} {\bibinfo {author} {\bibfnamefont {B.}~\bibnamefont
  {Wielinga}}\ and\ \bibinfo {author} {\bibfnamefont {G.~J.}\ \bibnamefont
  {Milburn}},\ }\href {\doibase 10.1103/PhysRevA.48.2494} {\bibfield  {journal}
  {\bibinfo  {journal} {Phys. Rev. A}\ }\textbf {\bibinfo {volume} {48}},\
  \bibinfo {pages} {2494} (\bibinfo {year} {1993})}\BibitemShut {NoStop}%
\bibitem [{\citenamefont {Boutin}\ \emph {et~al.}(2017)\citenamefont {Boutin},
  \citenamefont {Toyli}, \citenamefont {Venkatramani}, \citenamefont {Eddins},
  \citenamefont {Siddiqi},\ and\ \citenamefont {Blais}}]{Boutin17}%
  \BibitemOpen
  \bibfield  {author} {\bibinfo {author} {\bibfnamefont {S.}~\bibnamefont
  {Boutin}}, \bibinfo {author} {\bibfnamefont {D.~M.}\ \bibnamefont {Toyli}},
  \bibinfo {author} {\bibfnamefont {A.~V.}\ \bibnamefont {Venkatramani}},
  \bibinfo {author} {\bibfnamefont {A.~W.}\ \bibnamefont {Eddins}}, \bibinfo
  {author} {\bibfnamefont {I.}~\bibnamefont {Siddiqi}}, \ and\ \bibinfo
  {author} {\bibfnamefont {A.}~\bibnamefont {Blais}},\ }\href {\doibase
  10.1103/PhysRevApplied.8.054030} {\bibfield  {journal} {\bibinfo  {journal}
  {Phys. Rev. Appl.}\ }\textbf {\bibinfo {volume} {8}},\ \bibinfo {pages}
  {054030} (\bibinfo {year} {2017})}\BibitemShut {NoStop}%
\bibitem [{\citenamefont {Cochrane}\ \emph {et~al.}(1999)\citenamefont
  {Cochrane}, \citenamefont {Milburn},\ and\ \citenamefont
  {Munro}}]{CochranePRA99}%
  \BibitemOpen
  \bibfield  {author} {\bibinfo {author} {\bibfnamefont {P.~T.}\ \bibnamefont
  {Cochrane}}, \bibinfo {author} {\bibfnamefont {G.~J.}\ \bibnamefont
  {Milburn}}, \ and\ \bibinfo {author} {\bibfnamefont {W.~J.}\ \bibnamefont
  {Munro}},\ }\href {\doibase 10.1103/PhysRevA.59.2631} {\bibfield  {journal}
  {\bibinfo  {journal} {Phys. Rev. A}\ }\textbf {\bibinfo {volume} {59}},\
  \bibinfo {pages} {2631} (\bibinfo {year} {1999})}\BibitemShut {NoStop}%
\bibitem [{\citenamefont {Goto}(2016)}]{Goto16}%
  \BibitemOpen
  \bibfield  {author} {\bibinfo {author} {\bibfnamefont {H.}~\bibnamefont
  {Goto}},\ }\href {\doibase 10.1038/srep21686} {\bibfield  {journal} {\bibinfo
   {journal} {Scientific Reports}\ }\textbf {\bibinfo {volume} {6}},\ \bibinfo
  {pages} {21686} (\bibinfo {year} {2016})}\BibitemShut {NoStop}%
\bibitem [{\citenamefont {Puri}\ \emph {et~al.}(2017)\citenamefont {Puri},
  \citenamefont {Boutin},\ and\ \citenamefont {Blais}}]{Puri17}%
  \BibitemOpen
  \bibfield  {author} {\bibinfo {author} {\bibfnamefont {S.}~\bibnamefont
  {Puri}}, \bibinfo {author} {\bibfnamefont {S.}~\bibnamefont {Boutin}}, \ and\
  \bibinfo {author} {\bibfnamefont {A.}~\bibnamefont {Blais}},\ }\href
  {\doibase 10.1038/s41534-017-0019-1} {\bibfield  {journal} {\bibinfo
  {journal} {npj Quantum Information}\ }\textbf {\bibinfo {volume} {3}},\
  \bibinfo {pages} {18} (\bibinfo {year} {2017})}\BibitemShut {NoStop}%
\bibitem [{\citenamefont {Leghtas}\ \emph {et~al.}(2015)\citenamefont
  {Leghtas}, \citenamefont {Touzard}, \citenamefont {Pop}, \citenamefont {Kou},
  \citenamefont {Vlastakis}, \citenamefont {Petrenko}, \citenamefont {Sliwa},
  \citenamefont {Narla}, \citenamefont {Shankar}, \citenamefont {Hatridge},
  \citenamefont {Reagor}, \citenamefont {Frunzio}, \citenamefont {Schoelkopf},
  \citenamefont {Mirrahimi},\ and\ \citenamefont {Devoret}}]{ZakisSCI15}%
  \BibitemOpen
  \bibfield  {author} {\bibinfo {author} {\bibfnamefont {Z.}~\bibnamefont
  {Leghtas}}, \bibinfo {author} {\bibfnamefont {S.}~\bibnamefont {Touzard}},
  \bibinfo {author} {\bibfnamefont {I.~M.}\ \bibnamefont {Pop}}, \bibinfo
  {author} {\bibfnamefont {A.}~\bibnamefont {Kou}}, \bibinfo {author}
  {\bibfnamefont {B.}~\bibnamefont {Vlastakis}}, \bibinfo {author}
  {\bibfnamefont {A.}~\bibnamefont {Petrenko}}, \bibinfo {author}
  {\bibfnamefont {K.~M.}\ \bibnamefont {Sliwa}}, \bibinfo {author}
  {\bibfnamefont {A.}~\bibnamefont {Narla}}, \bibinfo {author} {\bibfnamefont
  {S.}~\bibnamefont {Shankar}}, \bibinfo {author} {\bibfnamefont {M.~J.}\
  \bibnamefont {Hatridge}}, \bibinfo {author} {\bibfnamefont {M.}~\bibnamefont
  {Reagor}}, \bibinfo {author} {\bibfnamefont {L.}~\bibnamefont {Frunzio}},
  \bibinfo {author} {\bibfnamefont {R.~J.}\ \bibnamefont {Schoelkopf}},
  \bibinfo {author} {\bibfnamefont {M.}~\bibnamefont {Mirrahimi}}, \ and\
  \bibinfo {author} {\bibfnamefont {M.~H.}\ \bibnamefont {Devoret}},\ }\href
  {\doibase 10.1126/science.aaa2085} {\bibfield  {journal} {\bibinfo  {journal}
  {Science}\ }\textbf {\bibinfo {volume} {347}},\ \bibinfo {pages} {853}
  (\bibinfo {year} {2015})}\BibitemShut {NoStop}%
\bibitem [{\citenamefont {Grimm}\ \emph {et~al.}(2020)\citenamefont {Grimm},
  \citenamefont {Frattini}, \citenamefont {Puri}, \citenamefont {Mundhada},
  \citenamefont {Touzard}, \citenamefont {Mirrahimi}, \citenamefont {Girvin},
  \citenamefont {Shankar},\ and\ \citenamefont {Devoret}}]{Grimm2020}%
  \BibitemOpen
  \bibfield  {author} {\bibinfo {author} {\bibfnamefont {A.}~\bibnamefont
  {Grimm}}, \bibinfo {author} {\bibfnamefont {N.~E.}\ \bibnamefont {Frattini}},
  \bibinfo {author} {\bibfnamefont {S.}~\bibnamefont {Puri}}, \bibinfo {author}
  {\bibfnamefont {S.~O.}\ \bibnamefont {Mundhada}}, \bibinfo {author}
  {\bibfnamefont {S.}~\bibnamefont {Touzard}}, \bibinfo {author} {\bibfnamefont
  {M.}~\bibnamefont {Mirrahimi}}, \bibinfo {author} {\bibfnamefont {S.~M.}\
  \bibnamefont {Girvin}}, \bibinfo {author} {\bibfnamefont {S.}~\bibnamefont
  {Shankar}}, \ and\ \bibinfo {author} {\bibfnamefont {M.~H.}\ \bibnamefont
  {Devoret}},\ }\href {https://www.nature.com/articles/s41586-020-2587-z}
  {\bibfield  {journal} {\bibinfo  {journal} {Nature}\ }\textbf {\bibinfo
  {volume} {584}},\ \bibinfo {pages} {205} (\bibinfo {year}
  {2020})}\BibitemShut {NoStop}%
\bibitem [{\citenamefont {Frattini}\ \emph {et~al.}(2022)\citenamefont
  {Frattini}, \citenamefont {Cortinas}, \citenamefont {Venkatraman},
  \citenamefont {Xiao}, \citenamefont {Su}, \citenamefont {Lei}, \citenamefont
  {Chapman}, \citenamefont {Joshi}, \citenamefont {Girvin}, \citenamefont
  {Schoelkopf}, \citenamefont {Puri},\ and\ \citenamefont
  {Devoret}}]{frattini2022squeezed}%
  \BibitemOpen
  \bibfield  {author} {\bibinfo {author} {\bibfnamefont {N.~E.}\ \bibnamefont
  {Frattini}}, \bibinfo {author} {\bibfnamefont {R.~G.}\ \bibnamefont
  {Cortinas}}, \bibinfo {author} {\bibfnamefont {J.}~\bibnamefont
  {Venkatraman}}, \bibinfo {author} {\bibfnamefont {X.}~\bibnamefont {Xiao}},
  \bibinfo {author} {\bibfnamefont {Q.}~\bibnamefont {Su}}, \bibinfo {author}
  {\bibfnamefont {C.~U.}\ \bibnamefont {Lei}}, \bibinfo {author} {\bibfnamefont
  {B.~J.}\ \bibnamefont {Chapman}}, \bibinfo {author} {\bibfnamefont {V.~R.}\
  \bibnamefont {Joshi}}, \bibinfo {author} {\bibfnamefont {S.~M.}\ \bibnamefont
  {Girvin}}, \bibinfo {author} {\bibfnamefont {R.~J.}\ \bibnamefont
  {Schoelkopf}}, \bibinfo {author} {\bibfnamefont {S.}~\bibnamefont {Puri}}, \
  and\ \bibinfo {author} {\bibfnamefont {M.~H.}\ \bibnamefont {Devoret}},\
  }\href@noop {} {\enquote {\bibinfo {title} {The squeezed kerr oscillator:
  spectral kissing and phase-flip robustness},}\ } (\bibinfo {year} {2022}),\
  \Eprint {http://arxiv.org/abs/2209.03934} {arXiv:2209.03934 [quant-ph]}
  \BibitemShut {NoStop}%
\bibitem [{\citenamefont {Iyama}\ \emph {et~al.}(2023)\citenamefont {Iyama},
  \citenamefont {Kamiya}, \citenamefont {Fujii}, \citenamefont {Mukai},
  \citenamefont {Zhou}, \citenamefont {Nagase}, \citenamefont {Tomonaga},
  \citenamefont {Wang}, \citenamefont {Xue}, \citenamefont {Watabe},
  \citenamefont {Kwon},\ and\ \citenamefont {Tsai}}]{Iyama23}%
  \BibitemOpen
  \bibfield  {author} {\bibinfo {author} {\bibfnamefont {D.}~\bibnamefont
  {Iyama}}, \bibinfo {author} {\bibfnamefont {T.}~\bibnamefont {Kamiya}},
  \bibinfo {author} {\bibfnamefont {S.}~\bibnamefont {Fujii}}, \bibinfo
  {author} {\bibfnamefont {H.}~\bibnamefont {Mukai}}, \bibinfo {author}
  {\bibfnamefont {Y.}~\bibnamefont {Zhou}}, \bibinfo {author} {\bibfnamefont
  {T.}~\bibnamefont {Nagase}}, \bibinfo {author} {\bibfnamefont
  {A.}~\bibnamefont {Tomonaga}}, \bibinfo {author} {\bibfnamefont
  {R.}~\bibnamefont {Wang}}, \bibinfo {author} {\bibfnamefont {J.-J.}\
  \bibnamefont {Xue}}, \bibinfo {author} {\bibfnamefont {S.}~\bibnamefont
  {Watabe}}, \bibinfo {author} {\bibfnamefont {S.}~\bibnamefont {Kwon}}, \ and\
  \bibinfo {author} {\bibfnamefont {J.-S.}\ \bibnamefont {Tsai}},\ }\href@noop
  {} {\enquote {\bibinfo {title} {Observation and manipulation of quantum
  interference in a superconducting kerr parametric oscillator},}\ } (\bibinfo
  {year} {2023}),\ \Eprint {http://arxiv.org/abs/2306.12299} {arXiv:2306.12299
  [quant-ph]} \BibitemShut {NoStop}%
\bibitem [{\citenamefont {Chou}\ \emph {et~al.}(2018)\citenamefont {Chou},
  \citenamefont {Blumoff}, \citenamefont {Wang}, \citenamefont {Reinhold},
  \citenamefont {Axline}, \citenamefont {Gao}, \citenamefont {Frunzio},
  \citenamefont {Devoret}, \citenamefont {Jiang},\ and\ \citenamefont
  {Schoelkopf}}]{Chou2018}%
  \BibitemOpen
  \bibfield  {author} {\bibinfo {author} {\bibfnamefont {K.~S.}\ \bibnamefont
  {Chou}}, \bibinfo {author} {\bibfnamefont {J.~Z.}\ \bibnamefont {Blumoff}},
  \bibinfo {author} {\bibfnamefont {C.~S.}\ \bibnamefont {Wang}}, \bibinfo
  {author} {\bibfnamefont {P.~C.}\ \bibnamefont {Reinhold}}, \bibinfo {author}
  {\bibfnamefont {C.~J.}\ \bibnamefont {Axline}}, \bibinfo {author}
  {\bibfnamefont {Y.~Y.}\ \bibnamefont {Gao}}, \bibinfo {author} {\bibfnamefont
  {L.}~\bibnamefont {Frunzio}}, \bibinfo {author} {\bibfnamefont {M.~H.}\
  \bibnamefont {Devoret}}, \bibinfo {author} {\bibfnamefont {L.}~\bibnamefont
  {Jiang}}, \ and\ \bibinfo {author} {\bibfnamefont {R.~J.}\ \bibnamefont
  {Schoelkopf}},\ }\href {\doibase 10.1038/s41586-018-0470-y} {\bibfield
  {journal} {\bibinfo  {journal} {Nature}\ }\textbf {\bibinfo {volume} {561}},\
  \bibinfo {pages} {368} (\bibinfo {year} {2018})}\BibitemShut {NoStop}%
\bibitem [{\citenamefont {Lifshitz}\ and\ \citenamefont
  {Cross}(2008)}]{LifshitzBook}%
  \BibitemOpen
  \bibfield  {author} {\bibinfo {author} {\bibfnamefont {R.}~\bibnamefont
  {Lifshitz}}\ and\ \bibinfo {author} {\bibfnamefont {M.~C.}\ \bibnamefont
  {Cross}},\ }\enquote {\bibinfo {title} {Nonlinear dynamics of nanomechanical
  and micromechanical resonators},}\ in\ \href {\doibase
  https://doi.org/10.1002/9783527626359.ch1} {\emph {\bibinfo {booktitle}
  {Reviews of Nonlinear Dynamics and Complexity}}}\ (\bibinfo  {publisher}
  {John Wiley \& Sons, Ltd},\ \bibinfo {year} {2008})\ Chap.~\bibinfo {chapter}
  {1}, pp.\ \bibinfo {pages} {1--52}\BibitemShut {NoStop}%
\bibitem [{\citenamefont {Dykman}\ \emph {et~al.}(1998)\citenamefont {Dykman},
  \citenamefont {Maloney}, \citenamefont {Smelyanskiy},\ and\ \citenamefont
  {Silverstein}}]{DykmanPRE98}%
  \BibitemOpen
  \bibfield  {author} {\bibinfo {author} {\bibfnamefont {M.~I.}\ \bibnamefont
  {Dykman}}, \bibinfo {author} {\bibfnamefont {C.~M.}\ \bibnamefont {Maloney}},
  \bibinfo {author} {\bibfnamefont {V.~N.}\ \bibnamefont {Smelyanskiy}}, \ and\
  \bibinfo {author} {\bibfnamefont {M.}~\bibnamefont {Silverstein}},\ }\href
  {\doibase 10.1103/PhysRevE.57.5202} {\bibfield  {journal} {\bibinfo
  {journal} {Phys. Rev. E}\ }\textbf {\bibinfo {volume} {57}},\ \bibinfo
  {pages} {5202} (\bibinfo {year} {1998})}\BibitemShut {NoStop}%
\bibitem [{\citenamefont {Wustmann}\ and\ \citenamefont
  {Shumeiko}(2013)}]{WustmanPRB13}%
  \BibitemOpen
  \bibfield  {author} {\bibinfo {author} {\bibfnamefont {W.}~\bibnamefont
  {Wustmann}}\ and\ \bibinfo {author} {\bibfnamefont {V.}~\bibnamefont
  {Shumeiko}},\ }\href {\doibase 10.1103/PhysRevB.87.184501} {\bibfield
  {journal} {\bibinfo  {journal} {Phys. Rev. B}\ }\textbf {\bibinfo {volume}
  {87}},\ \bibinfo {pages} {184501} (\bibinfo {year} {2013})}\BibitemShut
  {NoStop}%
\bibitem [{\citenamefont {Venkatraman}\ \emph {et~al.}(2023)\citenamefont
  {Venkatraman}, \citenamefont {Cortinas}, \citenamefont {Frattini},
  \citenamefont {Xiao},\ and\ \citenamefont {Devoret}}]{Venkatraman}%
  \BibitemOpen
  \bibfield  {author} {\bibinfo {author} {\bibfnamefont {J.}~\bibnamefont
  {Venkatraman}}, \bibinfo {author} {\bibfnamefont {R.~G.}\ \bibnamefont
  {Cortinas}}, \bibinfo {author} {\bibfnamefont {N.~E.}\ \bibnamefont
  {Frattini}}, \bibinfo {author} {\bibfnamefont {X.}~\bibnamefont {Xiao}}, \
  and\ \bibinfo {author} {\bibfnamefont {M.~H.}\ \bibnamefont {Devoret}},\
  }\href@noop {} {\enquote {\bibinfo {title} {A driven quantum superconducting
  circuit with multiple tunable degeneracies},}\ } (\bibinfo {year} {2023}),\
  \Eprint {http://arxiv.org/abs/2211.04605} {arXiv:2211.04605 [quant-ph]}
  \BibitemShut {NoStop}%
\bibitem [{\citenamefont {Eichler}\ and\ \citenamefont
  {Zilberberg}(2023)}]{EichlerBook23}%
  \BibitemOpen
  \bibfield  {author} {\bibinfo {author} {\bibfnamefont {A.}~\bibnamefont
  {Eichler}}\ and\ \bibinfo {author} {\bibfnamefont {O.}~\bibnamefont
  {Zilberberg}},\ }\href {https://books.google.ch/books?id=FZrWEAAAQBAJ} {\emph
  {\bibinfo {title} {Classical and Quantum Parametric Phenomena}}},\ Oxford
  Graduate Texts\ (\bibinfo  {publisher} {Oxford University Press},\ \bibinfo
  {year} {2023})\BibitemShut {NoStop}%
\bibitem [{\citenamefont {Kenfack}\ and\ \citenamefont
  {Zyczkowski}(2004)}]{KenfackJOB2004}%
  \BibitemOpen
  \bibfield  {author} {\bibinfo {author} {\bibfnamefont {A.}~\bibnamefont
  {Kenfack}}\ and\ \bibinfo {author} {\bibfnamefont {K.}~\bibnamefont
  {Zyczkowski}},\ }\href {\doibase 10.1088/1464-4266/6/10/003} {\bibfield
  {journal} {\bibinfo  {journal} {J. Opt. B: Quantum Semiclass. Opt.}\ }\textbf
  {\bibinfo {volume} {6}},\ \bibinfo {pages} {396} (\bibinfo {year}
  {2004})}\BibitemShut {NoStop}%
\bibitem [{\citenamefont {Walschaers}(2021)}]{MattiaPRXQ}%
  \BibitemOpen
  \bibfield  {author} {\bibinfo {author} {\bibfnamefont {M.}~\bibnamefont
  {Walschaers}},\ }\href {\doibase 10.1103/PRXQuantum.2.030204} {\bibfield
  {journal} {\bibinfo  {journal} {PRX Quantum}\ }\textbf {\bibinfo {volume}
  {2}},\ \bibinfo {pages} {030204} (\bibinfo {year} {2021})}\BibitemShut
  {NoStop}%
\bibitem [{\citenamefont {Spagnolo}\ \emph {et~al.}(2023)\citenamefont
  {Spagnolo}, \citenamefont {Brod}, \citenamefont {Galv{\~a}o},\ and\
  \citenamefont {Sciarrino}}]{nlBSamp2023}%
  \BibitemOpen
  \bibfield  {author} {\bibinfo {author} {\bibfnamefont {N.}~\bibnamefont
  {Spagnolo}}, \bibinfo {author} {\bibfnamefont {D.~J.}\ \bibnamefont {Brod}},
  \bibinfo {author} {\bibfnamefont {E.~F.}\ \bibnamefont {Galv{\~a}o}}, \ and\
  \bibinfo {author} {\bibfnamefont {F.}~\bibnamefont {Sciarrino}},\ }\href
  {\doibase 10.1038/s41534-023-00676-x} {\bibfield  {journal} {\bibinfo
  {journal} {npj Quantum Information}\ }\textbf {\bibinfo {volume} {9}},\
  \bibinfo {pages} {3} (\bibinfo {year} {2023})}\BibitemShut {NoStop}%
\bibitem [{\citenamefont {Paris}\ \emph {et~al.}(2003)\citenamefont {Paris},
  \citenamefont {Illuminati}, \citenamefont {Serafini},\ and\ \citenamefont
  {De~Siena}}]{Paris03}%
  \BibitemOpen
  \bibfield  {author} {\bibinfo {author} {\bibfnamefont {M.~G.~A.}\
  \bibnamefont {Paris}}, \bibinfo {author} {\bibfnamefont {F.}~\bibnamefont
  {Illuminati}}, \bibinfo {author} {\bibfnamefont {A.}~\bibnamefont
  {Serafini}}, \ and\ \bibinfo {author} {\bibfnamefont {S.}~\bibnamefont
  {De~Siena}},\ }\href {\doibase 10.1103/PhysRevA.68.012314} {\bibfield
  {journal} {\bibinfo  {journal} {Phys. Rev. A}\ }\textbf {\bibinfo {volume}
  {68}},\ \bibinfo {pages} {012314} (\bibinfo {year} {2003})}\BibitemShut
  {NoStop}%
\bibitem [{\citenamefont {Reagor}\ \emph {et~al.}(2016)\citenamefont {Reagor},
  \citenamefont {Pfaff}, \citenamefont {Axline}, \citenamefont {Heeres},
  \citenamefont {Ofek}, \citenamefont {Sliwa}, \citenamefont {Holland},
  \citenamefont {Wang}, \citenamefont {Blumoff}, \citenamefont {Chou},
  \citenamefont {Hatridge}, \citenamefont {Frunzio}, \citenamefont {Devoret},
  \citenamefont {Jiang},\ and\ \citenamefont {Schoelkopf}}]{ReagorPRB16}%
  \BibitemOpen
  \bibfield  {author} {\bibinfo {author} {\bibfnamefont {M.}~\bibnamefont
  {Reagor}}, \bibinfo {author} {\bibfnamefont {W.}~\bibnamefont {Pfaff}},
  \bibinfo {author} {\bibfnamefont {C.}~\bibnamefont {Axline}}, \bibinfo
  {author} {\bibfnamefont {R.~W.}\ \bibnamefont {Heeres}}, \bibinfo {author}
  {\bibfnamefont {N.}~\bibnamefont {Ofek}}, \bibinfo {author} {\bibfnamefont
  {K.}~\bibnamefont {Sliwa}}, \bibinfo {author} {\bibfnamefont
  {E.}~\bibnamefont {Holland}}, \bibinfo {author} {\bibfnamefont
  {C.}~\bibnamefont {Wang}}, \bibinfo {author} {\bibfnamefont {J.}~\bibnamefont
  {Blumoff}}, \bibinfo {author} {\bibfnamefont {K.}~\bibnamefont {Chou}},
  \bibinfo {author} {\bibfnamefont {M.~J.}\ \bibnamefont {Hatridge}}, \bibinfo
  {author} {\bibfnamefont {L.}~\bibnamefont {Frunzio}}, \bibinfo {author}
  {\bibfnamefont {M.~H.}\ \bibnamefont {Devoret}}, \bibinfo {author}
  {\bibfnamefont {L.}~\bibnamefont {Jiang}}, \ and\ \bibinfo {author}
  {\bibfnamefont {R.~J.}\ \bibnamefont {Schoelkopf}},\ }\href {\doibase
  10.1103/PhysRevB.94.014506} {\bibfield  {journal} {\bibinfo  {journal} {Phys.
  Rev. B}\ }\textbf {\bibinfo {volume} {94}},\ \bibinfo {pages} {014506}
  (\bibinfo {year} {2016})}\BibitemShut {NoStop}%
\bibitem [{\citenamefont {Steck}(2007)}]{steck2007quantum}%
  \BibitemOpen
  \bibfield  {author} {\bibinfo {author} {\bibfnamefont {D.~A.}\ \bibnamefont
  {Steck}},\ }\href
  {https://atomoptics.uoregon.edu/~dsteck/teaching/quantum-optics/quantum-optics-notes.pdf}
  {\bibfield  {journal} {\bibinfo  {journal} {Quantum and atom optics}\ }
  (\bibinfo {year} {2007})}\BibitemShut {NoStop}%
\end{thebibliography}%


\let\addcontentsline\oldaddcontentsline% Restore \addcontentsline

\clearpage
\newpage

\renewcommand{\thetable}{S\arabic{table}}  
\renewcommand{\thepage}{S\arabic{page}}  
\renewcommand{\thefigure}{S\arabic{figure}}
\renewcommand{\theHfigure}{S\arabic{figure}}
\renewcommand{\theequation}{S\arabic{equation}}
\setcounter{page}{1}
\setcounter{figure}{0}
\setcounter{table}{0}
\setcounter{section}{0}
\setcounter{equation}{0}

%\resetlinenumber
\widetext

{\centering\textbf{\Large Supplementary Material for} \\} 
{\centering\textbf{\Large Quantum squeezing in a nonlinear mechanical oscillator}\\} 
\normalsize
\vspace{.3cm}
{\centering Stefano Marti$^\ast$, Uwe von L\"upke$^\ast$, Om Joshi, Yu Yang, Marius Bild, Andraz Omahen, Yiwen Chu, and Matteo Fadel\\
\textit{Department of Physics, ETH Z\"{u}rich, 8093 Zurich, Switzerland}\\
\textit{Quantum Center, ETH Z\"{u}rich, 8093 Z\"{u}rich, Switzerland}\\
}

\suppressfloats

\tableofcontents

\vspace{10mm}
\section{Device and experiment parameters}
\begin{table}[H]
\centering
\renewcommand{\arraystretch}{1.5}
\begin{tabular}{lr|lr}
\hline
parameter & value & parameter & value \\
\hline
$\omega_{\text{qubit}}/2\pi$ &  5.042\,GHz &$\omega_{\text{phonon}}$ & 5.023\,GHz \\
$g_0/2\pi$& 229\,kHz &$\Delta_{\text{dispersive}}/2\pi$ & 1.85\,MHz\\
$\gamma_1/2\pi$ &  9.4\,kHz & $\kappa_1/2\pi$ & 1.2\,kHz \\
$\gamma_2^{\text{Ramsey}}/2\pi$ &  6.6\,kHz & $\kappa_2^{\text{Ramsey}}/2\pi$ & 0.76\,kHz \\
$\alpha/2\pi$ &  185\,MHz &  $\D_{21}/2\pi $ & 30\,MHz \\
\hline
\end{tabular}
\caption{\textbf{List of qubit and phonon properties and experimental parameters.} $\Delta_{\text{dispersive}}$ is the detuning between the qubit and phonon mode for the dispersive interaction used during Wigner function measurements.
$\gamma_1$ ($\kappa_1$) is the qubit (phonon) energy relaxation rate, and $\gamma_2^{\text{Ramsey}}$ ($\kappa_2^{\text{Ramsey}}$) is the qubit (phonon) decoherence rate measured by a Ramsey sequence. $\alpha$ is the qubit anharmonicity.}
\label{tab:device}
\end{table}

\clearpage
\newpage

\section{Deriving the Hamiltonian}\label{sec:squeezingRateDerivation}
In this section we derive the effective squeezing rate $g_{sq}$. 
We start from the Hamiltonian of a qubit with frequency $\oq$ and anharmonicity $-\alpha$ coupled to a single phonon mode with frequency $\oa$, and driven with two microwave drives at frequencies $\om_{1,2}$
\begin{equation}
    H_\mr{sys}/\hbar = \oq \qdq -\frac{\alpha}{2} \qd^2 q^2 + \oa \ada + g (\ad q + \hc) + \left(\Om_1 e^{-i\om_1 t} + \Om_2 e^{-i\om_2 t - i\phi} \right)\qd + \hc \, , \label{eq:Hsys}
\end{equation}
where $g$ is the qubit-phonon coupling strength, $\Om_{1,2}\in \mathcal{R}$ are the drive amplitudes, and $\phi$ is the initial phase difference between the drives. In the following, we take $\hbar=1$ for convenience. \\
We now enter a rotating frame at the qubit and phonon frequencies
\begin{equation}
   U_\mr{rf} = \exp{\left[i(\oq \qdq + \oa \ada )t\right]}\, , \label{eq:UrotFrame}
\end{equation}
in which the system Hamiltonian reads
\begin{equation}
   H_\mr{rf} = -\frac{\alpha}{2} \qd^2 q^2 + g (\ad q e^{i \D_a^{(0)} t} + \hc) + \left(\Om_1 e^{-i\D_1 t} + \Om_2 e^{-i\D_2 t - i\phi} \right)\qd + \hc \, , \label{eq:H_rf}
\end{equation}
where $\D_a^{(0)} = \oa - \oq$. 
Next, we enter an interaction picture of the microwave drives with the transformation
\begin{equation}
    U_\mr{d} =  \exp{\left[ \xi_1 e^{i\D_1 t}q + \xi_2 e^{i\D_2 t + i\phi}q - \hc \right]}\, , \label{eq:U_drives}
\end{equation}
where $\xi_{1,2} = \Om_{1,2} / |\D_{1,2}|$ are the relative drive amplitudes. 
$U_\mr{d}$ transforms the qubit operator as 
\begin{equation}
    q' = U_\mr{d} q U_\mr{d}^\dagger = q + \xi_1 e^{-i\D_1 t}q + \xi_2 e^{-i\D_2 t-i\phi}\, .
\end{equation}
Applying $U_\mr{d}$ to $H_\mr{rf}$ and using a rotating wave approximation to drop fast-oscillating terms, we find 
\begin{subequations}
 \begin{align}
H_\mr{d} &= U_\mr{d} H_\mr{rf} U_\mr{d}^\dag + i \dot{U}_\mr{d} U_\mr{d}^\dag \label{eq:H_d}\\
 &= \underbrace{-\frac{\alpha}{2} \qd^2 q^2}_{H_\mr{Kerr}} +  \underbrace{g (\ad q e^{i \D_a t} + \hc)}_{\mr{qubit-phonon~coupling}}    \underbrace{- \alpha \xi_1 \xi_2 (\qd^2 e^{-i\Sig t - i \phi} + \hc)}_\mr{two~photon~qubit~drive} \,.
 % + \left(\Om_1 e^{-i\D_1 t} + \Om_2 e^{-i\D_2 t - i\phi} \right)\qd + \hc \, , \label{eq:Hrf}
 % \underbrace{-2\alpha \left(\xi_1^2 + \xi_2^2 \right) \qdq}_{\D_q^{ss}} 
 \end{align}
\end{subequations}
Here, we have defined $\Sig = \D_1 + \D_2 = \om_1 + \om_2 - 2\om_q$. 
In $H_\mr{d}$ we have absorbed an AC Stark shift of the qubit frequency caused by the two microwave drives into the detuning $\D_a = \D_a^{(0)}-\D_q^\mr{Stark~shift}$. 
Of the terms that emerge from the drive transformation we kept the two-photon qubit drive through the rotating wave approximation. This is justified when we set up the drive frequencies symmetrically around the phonon mode such that $\Sig \sim 2\D_a$. \\
Since the two-photon operator $\qd^2+\hc$ acts on more than our usual computational subspace of the first two qubit energy levels, we now enter a qubit-state dependent rotating frame, in which we eliminate $H_\mr{Kerr}$. The transformation 
\begin{equation}
    U_\mr{K} = \exp{\left[-i\frac{\alpha}{2}\qd^2 q^2 t\right]} \label{eq:U_K}
\end{equation}
transforms the qubit operator as 
\begin{equation}
    U_\mr{K} q U_\mr{K}^\dag = e^{i\alpha t\qdq} q \, , ~ U_\mr{K} \qd U_\mr{K}^\dag = \qd e^{-i\alpha t\qdq} \mr{,~and~} U_\mr{K} \qd^2 U_\mr{K}^\dag = \qd^2 e^{-i\alpha t (2\qdq+1)} \, .\label{eq:U_K_q}
\end{equation}
With this transformation $H_\mr{d}$ becomes
\begin{subequations}
 \begin{align}
H_\mr{K} &= U_\mr{K} H_\mr{d} U_\mr{K}^\dag + i \dot{U}_\mr{K} U_\mr{K}^\dag \label{eq:H_K} \\
 & = g (\ad e^{i\alpha t\qdq} q e^{i \D_a t} + \hc) - \alpha \xi_1 \xi_2 (\qd^2e^{-i\alpha t (2\qdq+1)-i\Sig t - i \phi} + \hc) \label{eq:qubit-two-photon}\,.
\end{align}
\end{subequations}
The resonance conditions in the phase exponents of Eq. (\ref{eq:H_K}) now take the qubit anharmonicity into account. \\
Writing $M = \Sig + \alpha (2\qdq + 1)$, we can move to eliminate the two-photon qubit drive with a modified displacement transformation
\begin{equation}
    U_\mr{sq} = e^{S_\mr{sq}} \equiv  \exp{\left[ \alpha \xi_1 \xi_2 \qd^2 M^{-1} e^{-i M t -i \phi} -\hc \right]}\label{eq:U_sq}\, .
\end{equation}
$U_\mr{sq}$ transforms the qubit operator like 
\begin{subequations}
 \begin{align}
 U_\mr{sq} q U_\mr{sq}^\dag & \approx q + [S_\mr{sq}, q]  \label{eq:Usq-q-Usq} \\
 & = q + \alpha \xi_1 \xi_2 e^{-i(\Sig + \alpha) t - i\phi} \left[ \qd^2 M^{-1} e^{-2i\alpha t\qdq}, q \right] \\ 
 % & = \alpha \xi_1 \xi_2 \left( e^{-i(\Sig + \alpha) t - i\phi} \left[\qd^2, q \right]M^{-1} e^{-2i\alpha t\qdq } +  \qd^2\left[  M^{-1} e^{-2i\alpha t\qdq}, q \right]\right) \\ 
 & \approx q+ \alpha \xi_1 \xi_2 e^{-i(\Sig + \alpha) t - i\phi} \left[\qd^2, q \right]M^{-1} e^{-2i\alpha t\qdq } \label{eq:Usq-approximation} \\
 & = q-2\alpha \xi_1 \xi_2 e^{-i(\Sig + \alpha) t - i\phi} \qd M^{-1} e^{-2i\alpha t\qdq } \label{eq:q-sq}\, ,
\end{align}
\end{subequations}
where in Eq. (\ref{eq:Usq-approximation}) we neglected the commutator $\left[ M^{-1} e^{-2i\alpha t\qdq }, q\right]$, because, as we will see later, it only produces far off-resonant terms related to higher qubit levels and thus does not significantly affect the dynamics of our experiment. \\
The transformed Hamiltonian then reads
\begin{subequations}
 \begin{align}
H_\mr{c} &= U_\mr{sq} H_\mr{K} U_\mr{sq}^\dag + i \dot{U}_\mr{sq} U_\mr{sq}^\dag \label{eq:H_c} \\
 & = g (\ad e^{i\alpha t\qdq} q e^{i \D_a t} + \hc) - 2\alpha \xi_1 \xi_2 g \ad e^{i\alpha t\qdq} \qd e^{-2i\alpha t \qdq} M^{-1} e^{-i(\Sig + \alpha) t - i\phi} e^{i\D_a t} + \hc \\ 
 & = g (\ad e^{i\alpha t\qdq} q e^{i \D_a t} + \hc) - 2\alpha \xi_1 \xi_2 g \ad \qd e^{-i\alpha t \qdq} M^{-1} e^{-i(\Sig - \D_a) t - i\phi} + \hc \,.
\end{align}
\end{subequations}
As before, the first term in $H_\mr{c}$ describes the qubit-phonon coupling. The new, second term describes a two-mode interaction involving the simultaneous creation or annihilation of excitations in qubit and phonon mode, which becomes resonant when $\Sig \approx \D_a$. 
While higher qubit states play a role for the prefactor of the two-photon qubit drive in Eq. (\ref{eq:qubit-two-photon}), they do not participate in the phonon squeezing term we are looking for. 
Therefore, we now undo the level-dependent rotating frame transformation $U_\mr{K}$, by applying its inverse. 
\begin{subequations}
 \begin{align}
H_\mr{c}' &= U_\mr{K}^\dag H_\mr{c} U_\mr{K} + i \frac{\partial }{\partial t}\left(U_\mr{K}^\dag\right) U_\mr{K} \label{eq:H_c-prime} \\
 & = \underbrace{g (\ad  q e^{i \D_a t} + \hc)}_\mr{qubit-phonon~coupling} - 2\alpha \xi_1 \xi_2 g \left(\ad \qd  M^{-1} e^{-i(\Sig - \D_a) t - i\phi} + \hc\right) \underbrace{- \frac{\alpha}{2}\qd^2 q^2}_{H_\mr{Kerr}} \,.
\end{align}
\end{subequations}
Note, that the qubit anharmonicity is still described by this Hamiltonian in form of the reappearing $H_\mr{Kerr}$, but we can now treat the qubit-phonon interaction separate from this anharmonicity by transforming and interpreting the first two terms.
We now eliminate the qubit-phonon coupling term via the standard time-dependent Schrieffer-Wolff transformation
\begin{equation}
    U_\mr{SW} = \exp{\left[\frac{g}{\D_a}\ad q e^{i\D_at}-\hc\right]}\, ,\label{eq:U-SW}
\end{equation}
resulting in 
\begin{subequations}
 \begin{align}
H_\mr{sq} &= U_\mr{SW} H_\mr{c}' U_\mr{SW}^\dag + i \dot{U}_\mr{SW} U_\mr{SW}^\dag \label{eq:H_sq} \\
 & = - 2\alpha \xi_1 \xi_2 g \left(\ad \qd  M^{-1} e^{-i(\Sig - \D_a) t - i\phi} + \hc\right)  + U_\mr{SW} H_\mr{Kerr} U_\mr{SW}^\dag \nonumber \\
 & + \underbrace{\frac{g^2}{\D_a}\left(\ada - \qdq\right) }_\mr{normal-mode~splitting} -2 \frac{g^2}{\D_a} \alpha \xi_1 \xi_2 M^{-1}\left[ \underbrace{\left( \ad^2 e^{-i(\Sig - 2\D_a)t - i\phi} + \hc \right)}_\mr{phonon~squeezing} - \left( \qd^2 e^{-i\Sig t - i\phi} + \hc \right)\right]\, \label{eq:H_sq_explicit}.
\end{align}
\end{subequations}
In a final step, we assume or qubit is initially in its ground state $\ket{g}$. 
This results in $M = \Sig + \alpha$, after which we can write the phonon dynamics from Eq. (\ref{eq:H_sq_explicit}) for $\Sig \approx 2\D_a$ as 
\begin{equation}
    H_\mr{ph} = \frac{g^2}{\D_a} \ada -2 \frac{g^2}{\D_a} \frac{\alpha \xi_1 \xi_2}{\Sig + \alpha} \left( \ad^2 e^{-i(\Sig - 2\D_a)t - i\phi} + \hc \right) \label{eq:H-phonon} \, .
\end{equation}
Assuming the qubit starts in its ground state also eliminates the commutator we neglected in Eq. (\ref{eq:Usq-approximation}). \\
Eq. (\ref{eq:H-phonon}) contains the phonon frequency shift due to the normal-mode splitting with the qubit and the phonon squeezing term. 
In addition, we can include the anharmonicity of the phonon mode, which it inherits from the qubit due to their hybridization and which is included in $U_\mr{SW}H_\mr{Kerr}U_\mr{SW}^\dag$ in Eq. (\ref{eq:H_sq_explicit}). We derive the value $K$ of this phonon anharmonicity in Section \ref{sec:phonon-mode-anharmonicity} via time-independent perturbation theory (see Eq. (\ref{eq:phonon-Kerr}). 
Furthermore, we enter a frame rotating at the resonance condition of the squeezing interaction $\Sig - 2\D_a$, such that $H_\mr{ph}$ assumes the form of the squeezed Kerr oscillator (reintroducing here $\hbar$)
\begin{equation}
    H_\mr{ph,K}/\hbar = -\D \ada - \left(\epsilon \ad^2 + \hc\right) - K\ad^2a^2 \label{eq:H_mechanical_Kerr}\, , 
\end{equation}
where 
\begin{subequations}
 \begin{align}
 \D &= \Sig/2 - \frac{g^2}{\D_a} - \D_a  =(\omega_1 + \omega_2 - 2\omega_a')/2 \\ 
 \epsilon &= 2\dfrac{g^2}{\D_a}\xi_1\xi_2 \dfrac{\alpha}{\Sig + \alpha}e^{-i\phi}\\
 K &= \frac{g^4}{\D_a^3}\,. 
 \end{align}
\end{subequations}
We find that, as expected, we can tune the squeezing angle by varying $\phi$. In addition, we recover the squeezing strength $|\epsilon|$ derived via Floquet and perturbation theory in \cite{zhang2019engineering}.
\clearpage
\newpage

\section{Squeezing with losses and dephasing}\label{sec:lossySqEvolution}

\subsection{Preparation of a squeezed state with decoherence}

We consider here the evolution under the Hamiltonian
\begin{equation}
    H/\hbar = \Delta \ad a + \epsilon (a^2+{\ad}^2) \;,
\end{equation}
under the presence of losses and dephasing. As we will show, this can be solved analytically in an exact way.

The evolution of an operator $O$ can be computed as
\begin{equation}\label{eq:supp_dOdt}
    \dfrac{d\avg{O}}{dt} = \dfrac{i}{\hbar}\left\langle [H,O] \right\rangle + \avg{\mathcal{D}[O]} \;,
\end{equation}
where $\mathcal{D}[O]=\sum_i \frac{1}{2}\left(L_i^\dagger[O,L_i]+[L_i^\dagger,O]L_i\right)$ is the dissipator associated with the jump operators $L_i$. Considering energy relaxation described by $L_1=\sqrt{\gamma}a$, and pure dephasing described by $L_2=\sqrt{2\gamma_\phi} n$, we have
\begin{equation}
    \mathcal{D}[O] = \dfrac{\gamma}{2} \left( \ad [O,a] + [\ad,O]a \right) + \gamma_\phi \left( n [O,n] + [n,O]n \right) \;.
\end{equation}
Here, $\gamma=1/T_1$ and $\gamma_\phi=1/T_\phi$ with $T_\phi=(1/T_2-1/(2T_1))^{-1}$, where $T_1$ is the phonon energy relaxation time and $T_2$ the phonon Ramsey decoherence time.

From this we obtain
\begin{subequations}\label{eq:supp_eom_aad}
\begin{align}
    \dfrac{d \avg{a}}{dt} &= - i \Delta \avg{a} - 2 i \epsilon \avg{\ad} - \dfrac{1}{2} (\gamma+2\gamma_\phi) \avg{a} \;, \\
    \dfrac{d \avg{\ad a}}{dt} &= -i 2 \epsilon \left(\avg{\ad^2}-\avg{a^2}\right) - \gamma \avg{\ad a} \;, \\
    \dfrac{d \avg{a a}}{dt} &= -i 2\Delta \avg{a^2} - i 2 \epsilon (1+2\avg{\ad a}) - (\gamma + 4 \gamma_\phi) \avg{a^2} \;, 
\end{align}
\end{subequations}
that can be solved analytically for any desired initial conditions. This allows us to write the covariance matrix for the $x$ and $p$ quadratures by using the fact that
\begin{subequations}
\begin{align}
    \avg{x} &= \dfrac{1}{\sqrt{2}}\left(\avg{a} + \avg{\ad} \right) \\
    \avg{p} &= \dfrac{1}{i\sqrt{2}}\left(\avg{a} - \avg{\ad} \right) \\
    \avg{x^2} &= \dfrac{1}{2}\left( 1 + 2 \avg{\ad a} + \avg{a^2} + \avg{\ad^2} \right) \\
    \avg{p^2} &= \dfrac{1}{2}\left(1 + 2 \avg{\ad a} - \avg{a^2} - \avg{\ad^2}  \right) \\
    \avg{xp+px} &= i \left( \avg{\ad^2}-\avg{a^2}\right) \;.
\end{align}
\end{subequations}
The two eigenvalues of the covariance matrix corresponds to the squeezing and antisqueezing variances. These are
\begin{align}
    V_\text{min}\equiv \min_{\theta} \var{x \cos{\theta}+p\sin{\theta}} = \dfrac{1}{2}\left( 1 + 2(\avg{\ad a}-\abs{\avg{a}}^2) - 2 \abs{\avg{a^2}-\avg{a}^2} \right) \;,\\
    V_\text{max}\equiv \max_{\theta} \var{x \cos{\theta}+p\sin{\theta}} = \dfrac{1}{2}\left( 1 + 2(\avg{\ad a}-\abs{\avg{a}}^2) + 2 \abs{\avg{a^2}-\avg{a}^2} \right) \;.
\end{align}
The set of equations just presented allows us to obtain analytic expressions for the time evolution of the squeezed and antisqueezed variances. In particular, we looked at solutions that start from the vacuum state, \ie $\avg{a}=\avg{\ad a}=\avg{aa}=0$ at $t=0$. As these expressions are particularly lengthy, we will not present them here. However, we will present the special case $\Delta=0$, $\gamma_\phi=0$, which gives
\begin{align}
    V_\text{min} = \dfrac{1}{2}\dfrac{\gamma +4 \epsilon  e^{-t (\gamma +4 \epsilon )}}{ (\gamma +4 \epsilon )} \;,\\
    V_\text{max} = \dfrac{1}{2}\dfrac{\gamma - 4 \epsilon  e^{-t (\gamma - 4 \epsilon )}}{(\gamma - 4 \epsilon )} \;.
\end{align}

\subsection{Free evolution of a squeezed state with decoherence}

Equations~\eqref{eq:supp_eom_aad} allows us to compute also how a squeezed state evolves in the presence of energy relaxation and dephasing. For this, we set $\Delta=\epsilon=0$, which corresponds to a free evolution, and solve analytically Eqs.~\eqref{eq:supp_eom_aad} taking as initial conditions an ideal squeezed state with minimum variance $V_0 = e^{-4 r}/2$. We obtain that the squeezing and antisqueezing variances evolve as
\begin{align}
    V_\text{min} &= \dfrac{1}{2} e^{-2 t (\gamma +2 \gamma_\phi )} \left(e^{t (\gamma +4 \gamma_\phi )} \left(e^{\gamma  t}+\cosh (4 r)-1\right)-e^{\gamma  t} \sinh (4 r)\right) \;,\\
   V_\text{max} &= \dfrac{1}{2} e^{-2 t (\gamma +2 \gamma_\phi )} \left(e^{t (\gamma +4 \gamma_\phi )} \left(e^{\gamma  t}+\cosh (4 r)-1\right)+e^{\gamma  t} \sinh (4 r)\right) \;.
\end{align}
If we consider $\gamma_\phi=0$, we obtain
\begin{equation}\label{eq:vDecFreeEv}
    V_\text{min} = \dfrac{1}{2} \left(1 + e^{-\gamma t}(2 V_0 - 1)\right) \;,
\end{equation}
which is used in the main text to fit the measurements shown in Fig.~1f.

\subsection{Thermal occupation of a Gaussian state}

Gaussian states (both pure and mixed) are fully determined by the vector of first moments and by their covariance matrix
\begin{equation}
    \Gamma = \begin{pmatrix}
\text{Var}[x] & \text{Cov}[x,p] \\
\text{Cov}[x,p] & \text{Var}[p]
\end{pmatrix} \;.
\end{equation}
The purity of a Gaussian state can then be expressed from $\Gamma$ as \cite{Paris03}
\begin{equation}
    \text{Tr}[\rho^2] = \dfrac{1}{2\sqrt{\text{Det}[\Gamma]}} = \dfrac{1}{1+2 n_T}\;,
\end{equation}
where $n_T$ is the mean number of thermal (\ie incoherent) excitations. In the basis that diagonalizes $\Gamma$, namely the one of the squeezed/antisqueezed quadratures, we thus have
\begin{equation}
    n_T = \sqrt{V_\text{min} V_\text{max}} - \dfrac{1}{2} \;.
\end{equation}
As expected, for a state that saturates the Heisenberg uncertainty relation we have $V_\text{min} V_\text{max} = 1/4$, and thus $n_T=0$.

\clearpage
\newpage
\section{Squeezing limits}\label{sec:SqLimit}

\subsection{Limits from energy relaxation and dephasing}

The off-resonant coupling to the transmon results in a modified energy relaxation and dephasing rate for the phonon. 
The effective energy relaxation rate for the phonon can be written as
\begin{equation}
    \kappa = \dfrac{1}{T_1^p} + \dfrac{g^2}{\Delta_a^2} \gamma \;,
\end{equation}
where $\gamma=1/T_1^q$ is the qubit energy relaxation rate. 

\begin{figure}[h!]
    \centering
    \includegraphics[width=\textwidth/2]{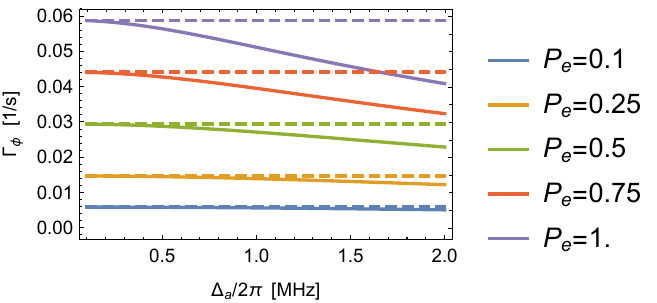}
    \caption{Phonon dephasing inherited from the qubit, see Eq.~\eqref{eq:inDep}.}
    \label{fig:suppDephasing}
\end{figure}

The dephasing rate can be approximated to be \cite{ReagorPRB16}
\begin{equation}\label{eq:inDep}
    \gamma_\phi = \gamma_\phi^0 + \Gamma_{\phi}(P_e,\Delta_a) \;,
\end{equation}
where $\gamma_\phi^0=(1/T_2^p-1/(2T_1^p))$ is the bare phonon dephasing rate, and
\begin{equation}
   \Gamma_{\phi}(P_e,\Delta_a) = \dfrac{\gamma}{2} \; \text{Re}\left[\sqrt{\left(1+\dfrac{2i \chi}{\gamma}\right)^2 + \dfrac{8 i \chi P_e}{\gamma}} - 1 \right] \;,
\end{equation}
with $\chi=2g^2/\Delta_a$ the dispersive shift and $P_e$ the qubit excited state population. For $g=2\pi\cdot\unit{0.29}{MHz}$ and $T_1^q=\unit{17}{\mu s}$ we plot in Fig.~\ref{fig:suppDephasing} the value of $\Gamma_{\phi}(P_e,\Delta_a)$ (solid lines) compared to the approximation $\Gamma_{\phi}(P_e,\Delta_a) \approx P_e \gamma$ valid when $\chi\gg\gamma$ (dashed lines).

\begin{figure}[h!]
    \centering
    \includegraphics[width=\textwidth/2]{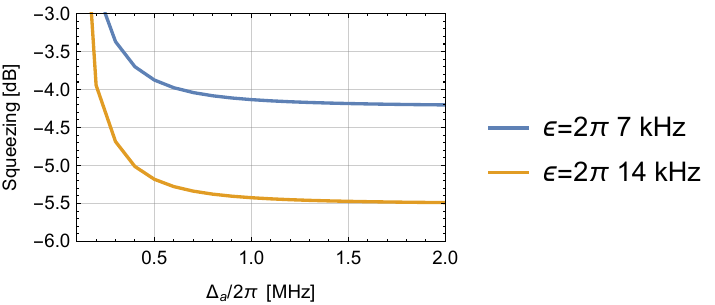}
    \caption{Maximum squeezing achievable at a given $\Delta_a$, for different values of $\epsilon$. }
    \label{fig:suppMaxSq}
\end{figure}

For a squeezing dynamics $H/\hbar=\epsilon({\ad}^2+a^2)$ in the presence of losses $\sqrt{\kappa}a$ and dephasing $\sqrt{2\gamma_\phi}\ad a$, the evolution of the squeezed variance as a function of time can be computed analytically (see Eq.~\eqref{eq:supp_eom_aad}). Therefore, for given values of the parameters we can look for the minimum variance as a function of time, and thus for the maximum squeezing achievable.
For $T_1^p=\unit{100}{\mu s}$, $T_2^p=\unit{150}{\mu s}$ that implies $\gamma_\phi^0=(\unit{600}{s})^{-1}$, and $P_e=0.1$ we obtain Fig.~\ref{fig:suppMaxSq}. Note however that, as we will see later, other effects will contribute in limiting the observed squeezing.

\clearpage
\newpage

\subsection{Limits from Kerr nonlinearity}
Even in the absence of energy relaxation and dephasing, the squeezing of a Kerr oscillator is going to be limited by the nonlinearity. While for short times the state evolves according to a simple squeezing dynamics, for longer times the Kerr nonlinearity results in a twisting of the state into an ``S'' shape. This implies an optimal evolution time $t^\ast$ at which the squeezing is maximized.

\begin{figure}[h!]
    \centering
    \includegraphics[width=\textwidth/2]{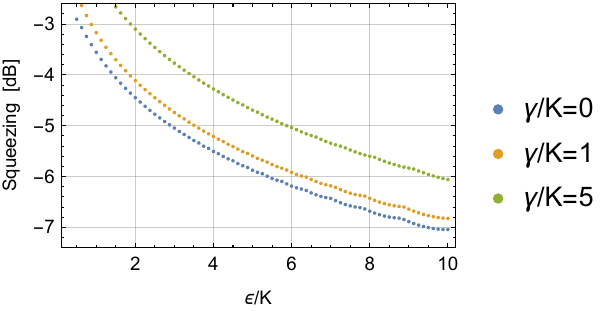}
    \caption{Maximum squeezing achievable for a given $\epsilon/K$, including energy relaxation at rate $\gamma$.}
    \label{fig:suppMaxSqKerr}
\end{figure}

Computing an analytical expression for $t^\ast$ for the squeezed Kerr Hamiltonian is challenging, especially if decoherence channels are considered. For this reason we perform a numerical simulation of how the vacuum state evolves under $H=\epsilon (a^2+{\ad}^2) - K \ad\ad a a$, considering energy relaxation at rate $\gamma$, and extract the maximum squeezing achievable for different values of the parameters. The results are shown in Fig.~\ref{fig:suppMaxSqKerr}. As an example, from the data presented in Fig.~2a we found $\epsilon=2\pi\cdot\unit{7.6}{kHz}$ and $\gamma=2\pi\cdot\unit{12.4}{kHz}$. Since $\Delta_a=2\pi\cdot\unit{1.5}{MHz}$, we expect from exact diagonalization $K=2\pi\cdot\unit{1.8}{kHz}$, which implies $\epsilon/K=4.2$ and $\gamma/K=6.8$. For these parameters, our simulation predicts a maximum squeezing of $\unit{-4.0}{dB}$, which is compatible with what we measure in Fig.~2a taking into account finite measurement time (see the next section).

\subsection{Limits from measurement time}
\begin{figure}[h!]
    \centering
    \includegraphics[width=\textwidth/2]{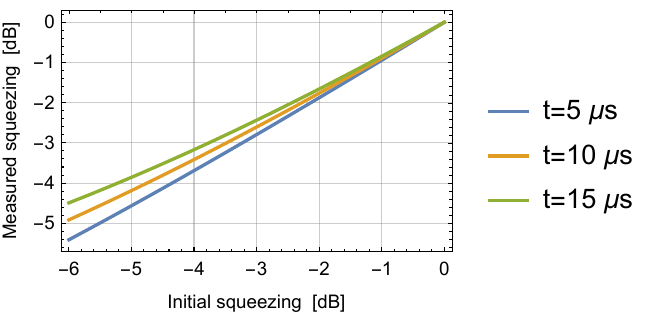}
    \caption{Squeezing loss due to finite measurement time $t$.}
    \label{fig:measSq}
\end{figure}

The measurement needed to characterize the squeezed state that has been prepared will inevitably take a finite time. In our case, this measurement is a Wigner function measurement, and it thus consists of a $\unit{5}{\mu s}$ long displacement pulse followed by a $\unit{5.7}{\mu s}$ long parity-echo sequence \cite{vonLupke22}. During this time, the state will decohere due to phonon relaxation and dephasing, to which the qubit also contributes due to its proximity in frequency when operating in the strong dispersive regime ($\Delta_a\approx 2\pi\cdot\unit{2}{MHz}$ during the parity measurement). This effectively results in an averaging of the measurement over a squeezed states decaying towards the vacuum. Since a precise modeling of this dynamics is complicated, to get an estimate for this effect we consider a simplified situation. We imagine that during the measurement time $t$ the state is only subject to energy relaxation, which result in a change of variance as described by Eq.~\eqref{eq:vDecFreeEv}. For $\gamma=\unit{100}{\mu s^{-1}}$, we show in Fig.~\ref{fig:measSq} the relation between the initial and the measured squeezing, for different values of $t$. As we did not take into account other effects taking place during the measurement time, such as dephasing, we take this analysis as a lower bound on the amount of squeezing that is lost during time $t$. From another point of view, this can be seen as a bound on the maximum measurement time acceptable in order to see a desired level of squeezing.

\clearpage
\newpage

\clearpage
\newpage
\section{Phonon mode anharmonicity}

\subsection{Anharmonicity from perturbation theory}\label{sec:phonon-mode-anharmonicity}
\hfill\\
Due to its interaction with the qubit, the phonon mode inherits an anharmonicity. To calculate the magnitude of this effect, let us start from the Hamiltonian of the system written in the rotating frame of the qubit
\begin{equation}\label{eq:fullJC}
    H = \underbrace{\Delta_a \ad a - \dfrac{\alpha}{2} \qd\qd q q}_{H_0} \; + \underbrace{\phantom{\dfrac{1}{2}} g (\qd a + q \ad)}_{\lambda V} \;.
\end{equation}
The energy of a state $\ket{n,l}$, where $n$ indicates the qubit state and $l$ the oscillator state, can be computed via perturbation theory in the parameter $\lambda$. To zeroth order we have
\begin{equation}
    E_{nl}^{(0)} = \left(\Delta_a l - \dfrac{\alpha}{2}n(n-1)\right) \ket{n,l} \;.
\end{equation}

\begin{figure}[h!]
    \centering
    \includegraphics[width=0.9\textwidth]{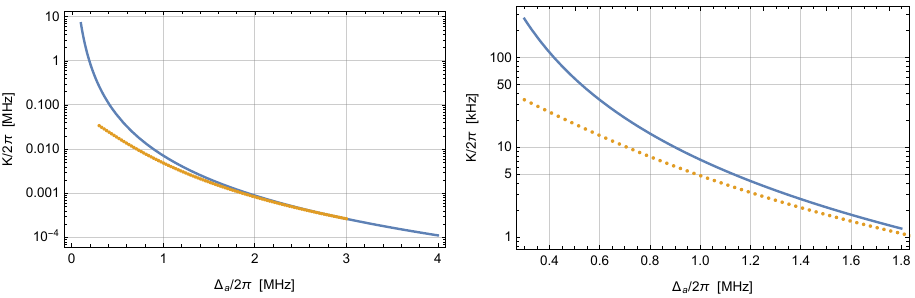}
    \caption{Phonon anharmonicity inherited from the qubit. Blue line is the result Eq.~\eqref{eq:phonon-Kerr} from perturbation theory, while yellow dots are the result obtained from numerical diagonalisation of the system Hamiltonian.}
    \label{fig:suppAnharm}
\end{figure}

Due to the form of $V$, it is easy to see that all odd-orders corrections $E_{n,l}^{2k+1}$ are zero. On the other hand, even-order corrections are in general non-zero. For the second order we find
\begin{equation}
    E_{nl}^{(2)} = g^2 \left( \dfrac{l(n+1)}{n\alpha+\Delta_a} - \dfrac{n(l+1)}{\alpha(n-1) + \Delta_a} \right) \;.
\end{equation}
The fourth order expression is lengthy, but if we restrict it to $n=0$ (\ie qubit in the ground state), we find
\begin{equation}
    E_{0l}^{(4)} = - \dfrac{g^4 l (l\alpha^2 + 2 l \alpha \Delta_a+(2l-1) \Delta_a^2)}{\Delta_a^3(\alpha+\Delta_a)^2} \;.
\end{equation}
From the above expressions we define $E_{l} \equiv E_{0l}^{(0)}+E_{0l}^{(2)}+E_{0l}^{(4)}$, from which we compute the phonon anharmonicity as
\begin{align}
    2 K &\equiv (E_1 - E_0) - (E_2 - E_1) \\
    &= \dfrac{2 g^4}{\Delta_a^3}\left( 1 + \dfrac{\Delta_a^2}{(\alpha+\Delta_a)^2} \right) \label{eq:phonon-Kerr}\;, 
\end{align}
that results in an effective Kerr Hamiltonian for the phonon
\begin{equation}
    H_K = - K \ad\ad a a \;.
\end{equation}
For our experimental parameters, $g=2\pi\cdot\unit{0.29}{MHz}$ and $\alpha=2\pi\cdot\unit{185}{kHz}$, we plot in Fig.~\eqref{fig:suppAnharm} the value of $K$ as a function of $\Delta_a$. For comparison, we also plot the value of $K$ obtained by numerical diagonalization of Eq.~\eqref{eq:fullJC}.

\clearpage
\newpage

\subsection{The steady-state of a driven Duffing-oscillator}
In Figure 3 of the main text we show measurements of the qubit response to a probe tone around the phonon frequency. 
We use these measurements to extract the anharmonicity of the phonon mode via the following fit:  
The steady state of the equation of motion of a Kerr oscillator with a classical amplitude $a(t)$ and frequency $\omega_a$ driven by a probe with frequency $\omega_p$ and amplitude $\Omega_p$ in the rotating frame of the probe is 
\begin{equation}
    a(t)\left[ i \left(\alpha_m |a(t)|^2 - \D_p\right) + \frac{\kappa}{2}\right] = \Omega_p \,. \label{eq:supp_eom_a}
\end{equation}
Here, $\D_p = \omega_p - \omega_a$ is the probe-phonon mode detuning, $\alpha_m = \om_a^{0\rightarrow 1} - \om_a^{1\rightarrow 2} = 2K$ is the anharmonicity of the phonon mode, and $\kappa$ is the phonon linewidth.
Multiplying Eq. (\ref{eq:supp_eom_a}) with its complex conjugate results in 
\begin{equation}
    \alpha_m^2 \bar{n}^3 - 2\D_p \alpha_m \bar{n}^2 + \left( \D_p^2 +\frac{\kappa^2}{4}\right)\bar{n} = \Omega_p^2\, , \label{eq:supp_eom_n}
\end{equation}
where $\bar{n}=|a(t)|^2$ is the average phonon occupation of the oscillator. 
Note that $\Om_p$ is the effective amplitude of the probe drive affecting the phonon mode, rather than the amplitude of the microwave drive we use to generate that probe tone. 
Furthermore, by assuming our phonon mode can be described by the classical equation of motion, Eq. (\ref{eq:supp_eom_a}), we ignore the dynamics that result from its quantization. 
We also neglect contributions of the qubit to the equation of motion of the phonon and assume we can treat the phonon mode by itself. \\
Before we can solve Eq. (\ref{eq:supp_eom_n}) for $\bar{n}$ and fit the solution to the measured data, we need to convert the measured qubit population to an inferred phonon population. 
To achieve this, we again assume we can treat the phonon mode classically, reducing its effect on the qubit to that of an off-resonant drive on a two-level system. 
Following Ref. \cite{steck2007quantum} (starting from Eq. (5.132) therein and using $\frac{1}{T_1^q}\approx\frac{1}{T_2^q}\ll\D_a$), we can write the steady-state population of an off-resonantly driven qubit as 
\begin{equation}
    P_e = \frac{2\Om_a^2}{\D_a^2 + 4\Om_a^2}\, .\label{eq:Pe_ss}
\end{equation}
Here, $\Om_a$ is the effective drive amplitude that the qubit experiences from the populated phonon mode. 
We can write $|\Om_a|^2 = g_a^2 \bar{n}$ and invert Eq. (\ref{eq:Pe_ss}) to arrive at an expression for the average phonon number $\bar{n}$
\begin{equation}
    \bar{n} = \frac{1}{g_a^2}\frac{P_e\D_a^2}{2-4P_e}\, .\label{eq:inferred_n}
\end{equation}
Using Eq. (\ref{eq:inferred_n}), we can now convert the measured qubit population into an inferred phonon population which we numerically fit the solution of Eq. (\ref{eq:supp_eom_n}) to. \\
\subsection{Fit routine for phonon anharmonicity measurements}
To fit the spectroscopy measurements, we first convert the measured qubit populations to average phonon numbers, using Eq. \ref{eq:inferred_n}. 
We then compute the solution of Eq. (\ref{eq:supp_eom_n}) with a set of initial parameters $\alpha_m$, $\kappa$, $\omega_a$, and $\Omega_p$. 
Next, we compute the absolute difference between this solution and the average phonon populations, which we use as residual function. 
To improve the fit quality, we use a weight that slightly prioritizes data points in the vicinity of the expected phonon-qubit detuning $\omega_a'$. 
In this first fit, we keep $\omega_a$ fixed to its initial value, resulting in approximate fitting parameters for $\alpha_m$, $\kappa$, and $\Omega_p$, which we use in a second fit as initial values, this time also fitting for $\omega_a$. 
This two-stage fitting procedure lets us narrow down better initial values, making the fit more robust across a wider range of measurements. 
From the covariance matrix of the second fit, we also extract error bars of the fit parameters. \\
% 

%\clearpage
%\newpage

\subsection{Onset of bistability}
Equation~\eqref{eq:supp_eom_n} is a cubic polynomial in $\bar{n}$, thus admitting either one or three real solutions. For sufficiently large $\Omega_p$, Eq.~\eqref{eq:supp_eom_n} admits three real solutions for $\Delta_p^- < \Delta_p < \Delta_p^+$, where $\Delta_p^\pm$ are the so-called saddle-node bifurcation points. These two are the real solution of $d \Delta_p/d \bar{n}=0$, and can thus be found from the differential form of Eq.~\eqref{eq:supp_eom_n}, 
\begin{equation}
    \left(3\alpha_m^2 \bar{n}^2 - 4 \Delta_p \alpha_m \bar{n} + \Delta_p^2 + \dfrac{\kappa^2}{4}\right) d\bar{n} = 2(\alpha_m \bar{n}^2 - \Delta_p \bar{n}) d\Delta_p \;.
\end{equation}
The solutions of $d \Delta_p/d \bar{n}=0$ are thus
\begin{equation}
    \Delta_p^\pm = 2\alpha_m\bar{n} \pm \dfrac{1}{2}\sqrt{4\alpha_m^2\bar{n}^2-\kappa^2} \;,
\end{equation}
which are real and distinct only for a sufficiently large probe amplitude $\Omega_p>\Omega_p^c$. The critical value $\Omega_p^c$ where the two bifurcation points emerge determines the onset of bistability, and it is given by having both $d \Delta_p/d \bar{n}=0$ and $d^2 \Delta_p/d \bar{n}^2=0$. At this critical point we have
\begin{equation}
    \bar{n}^c = \left\vert\dfrac{\kappa}{\sqrt{3} \alpha_m} \right\vert \;.
\end{equation}
For our values $\kappa=2\pi\cdot\unit{1.2}{kHz}$ and $\alpha_m=2K=2(2\pi\cdot\unit{14}{kHz})$ we obtain $\bar{n}^c=0.025$, namely $\vert a^c\vert=\sqrt{\bar{n}^c}=0.16$.

\vspace{10mm}
\section{Error bars calculations}\label{sec:errorbars}
\hfill\\

Errors on qubit and phonon $T_1$ and $T_2^\ast$ are one standard deviation (1STD) errors on the fit parameters, obtained from the square root of the diagonal elements of the fit covariance matrix. 
Errors on $V_\text{min}$ and $V_\text{max}$ extracted from 2D Gaussian fit of the Wigner functions are 95\% confidence intervals. 
Errors on $V_\text{min}$ and $V_\text{max}$ extracted from maximum likelihood state reconstruction are given by changing the Hilbert space truncation of $15$ by $\pm 2$, that we note to be the dominant error source (\ie bigger than imperfect axes calibration or Wigner background fluctuations).
Errors on the squeezing decay time, squeezing rate and effective decay time are 1STD errors on the fit parameters, that take into account the error on the individual fitted points.
Error on the Kerr nonlinearity are errors on the fit parameters.
Error on the quantum Fisher information are given by changing in the maximum likelihood state reconstruction algorithm the Hilbert space truncation of $15$ by $\pm 2$.

\end{document}